\documentclass[aps,prb,floatfix]{revtex4-2}
\usepackage{amssymb}
\usepackage{amsmath}
\usepackage{graphicx}
\usepackage{float}

\begin{document}

\title{{Time domain non-linear quantum interferometry}}
\author{ C. Fruy${}^{1,2,\textrm{a}}$, A. Th\'ery${}^{1,2,}$\footnote{These authors contributed equally to this work}, B. Hue${}^{1,2}$, W. Legrand${}^{3}$, L. Jarjat${}^{1,2}$, J. Craquelin${}^{1,2}$, M.R. Delbecq${}^{1,2,4}$, A. Cottet$^{1,2}$ and T. Kontos$^{1,2,5}$\footnote{To whom correspondence should be addressed: takis.kontos@ens.fr}}
\affiliation{${}^{1}$ Laboratoire de Physique de l'\'{E}cole Normale Sup\'{e}rieure, ENS, Universit\'{e} PSL, CNRS, Sorbonne Universit\'{e}, Universit\'{e} Paris-Diderot, Sorbonne Paris Cit\'{e}, Paris, France.}
\affiliation{$^{2}$Laboratoire de Physique et d'Etude des Mat\'eriaux, ESPCI Paris, PSL University, CNRS, Sorbonne Universit\'e, Paris, France}
\affiliation{$^{3}$Univ. Grenoble Alpes, CNRS, Grenoble INP, Institut N\'eel, 38000 Grenoble, France}
\affiliation{$^{4}$Institut universitaire de France (IUF)}
\affiliation{$^{5}$Institute of Astrophysics, FORTH, GR-71110 Heraklion, Greece}

\begin{abstract}
Interferometry is a powerful method for studying many fundamental phenomena ranging from gravitational waves to anyon excitations. Besides loss of coherence in the interferometer paths, quantum mechanics sets an incompressible bound for linear interferometry arising from the quantum noise of the light beam, called the shot noise limit. Here, we implement a non-linear quantum interferometer that uses a microwave cavity and a magnetic field resilient anharmonic superconducting quantum circuit made of granular aluminum. The circuit enables interferometry beyond the shot noise limit. A direct application of our findings is the detection of dark matter (axions or dark photons), high-frequency gravitational waves or astronomical masers. 

\end{abstract}
\date{\today}
\maketitle

\section{INTRODUCTION}

Interferometry is a well established technique for sensing classical and quantum signals\cite{review_int:15}. It can be developed both for electromagnetic waves and matter waves. It is, for example, at the heart of the detection of gravitational waves. It can also be used as a resource in condensed matter, for example for sensing the phase of the superconducting order parameter\cite{Jeon:23}. 

Improving the sensitivity of interferometry is an important and active generic question. One limitation of conventional interferometry is related to the noise of the light beam intensity arising from its quantumness, called the shot noise limit. There are two generic ways to mitigate this hurdle: first, one can seek to perform interferometry with a noise lower than the poissonian noise of a coherent state of light for example using squeezed light\cite{review_int:15}, second, one may use non-linear processes in order to introduce amplification \cite{QUMPI:23}. In both cases, anharmonicity of quantum circuits embedded in the paths of the light beams are a key resource for inducing the necessary photon correlations. 

Interestingly, it has been suggested recently to use a Kerr medium in order to enhance the phase signal of an interferometer, thereby surpassing the standard quantum limit of interferometry \cite{Kerrmedium:22}. The idea is to encode directly the phase information in the unitary evolution of the beam via the anharmonicity of the photon energy provided by the Kerr medium.  

Here, we implement a setup using a cavity quantum electrodynamics architecture comprising a microwave cavity and a granular aluminum Josephson \cite{Thery:24,Winkel:20,grAl:23} junction as a non-linear quantum circuit element. Thanks to the anharmonicity of the circuit, we can measure directly the interference between a cavity drive of large photon number and a test drive of very low photon number by performing time domain manipulations of our quantum circuit. Our non-linear quantum interferometer enables the detection of weak signals populating the cavity mode with a photon number as small as $10^{-2}$ in $400$~$\mu s$ and $10^{-5}$ in $400$~ms.

The immediate application we have in mind for our setup is precision measurements for cosmological signals. For example, many signals from the Dark Universe are expected to affect the electromagnetic spectrum in the microwave range. In terms of dark matter, the axions, hypothetical particles outside the Standard Model,\cite{PecceiQuinn:77, Weinberg:78,Wilczek:78} constitute an important candidate. They would modify profoundly Maxwell's equations by the addition of a magneto-electric term, proportional to the axion amplitude, oscillating at the axion mass\cite{review:1,review:2}. Since one of the most motivated range of masses is the microwave range, microwave amplification techniques have been known for long as effective tools to track axions, as epitomized by Sikivie's haloscope method\cite{Sikivie:83,Malnou:19,Crescini:20,Brubaker:17,Braine:20,Backes:21}. Similar effects on the electromagnetic field are expected for dark photons\cite{Dixit:19,DarkPhotons:21}, other candidates for dark matter. It is important to stress that, while the frequency/mass of axions is unknown, the axion paradigm has clear predictions in terms of the form and amplitude of their wave \cite{review:1,review:2}. It is a coherent field of amplitude directly linked to the density of dark matter at our location in the Milky Way and to the expected axion-photon coupling strength. The two well-established benchmark models are the KSVZ and DFSZ models \cite{review:1,review:2}. Another aspect well established of potential axion signals is their phase coherence time, in the $100 \mu s -ms$ range, arising from the velocity distribution in the galactic halo which leads to Doppler broadening of the linewidth of axions. This calls for the use of interferometric methods to probe these dark matter candidates. 

The haloscope technique can also be used to detect other objects in addition to dark matter, such as high-frequency gravitational waves\cite{Raffaele:22,RADES:24}. Other cosmic objects, less elusive, but still very important for astrophysics, the astronomical masers, can have similar effects with comparable powers (typically also of the order of $10^{-23}$ W) in Earth-based experiments, such as the OH maser at about $1.665$ GHz \cite{Maser:1965}. Hence, microwave quantum sensing techniques\cite{Clerk:10} are overall appealing to unravel deep questions about the Universe. A common feature of the above-mentioned signals is their narrow linewidth which arises from their weak Doppler broadening in space\cite{DarkPhotons:21,CAPP:22,Turner:90,Maser:1965}. The quality factor of the expected lines is typically around $10^6$, implying linewidths in the kHz range. Interferometry is therefore also an interesting method for detecting them. 

Interferometry has a major asset with respect to power measurements commonly used in haloscopes as one detects an amplitude and not a power which allows to gain "a square power" for sensing of a weak signal. This allows one to envision fast sensing  to deal with dephasing, the randomness of the initial phase being a simple offset which can be evaluated by repeating the sensing protocol only a few times. It is the aim of our paper to study how fast interferometry of weak photon signals can be performed in an environment compatible with haloscopes to give perspective for interferometric methods for dark matter quantum sensing. Note that the fact that the mass is unknown has similar consequences as all the existing detection strategies in terms of methodology\cite{review:1,review:2}. 

\begin{figure}[tph]
{\small \centering\includegraphics[width=0.65\linewidth,angle=0]{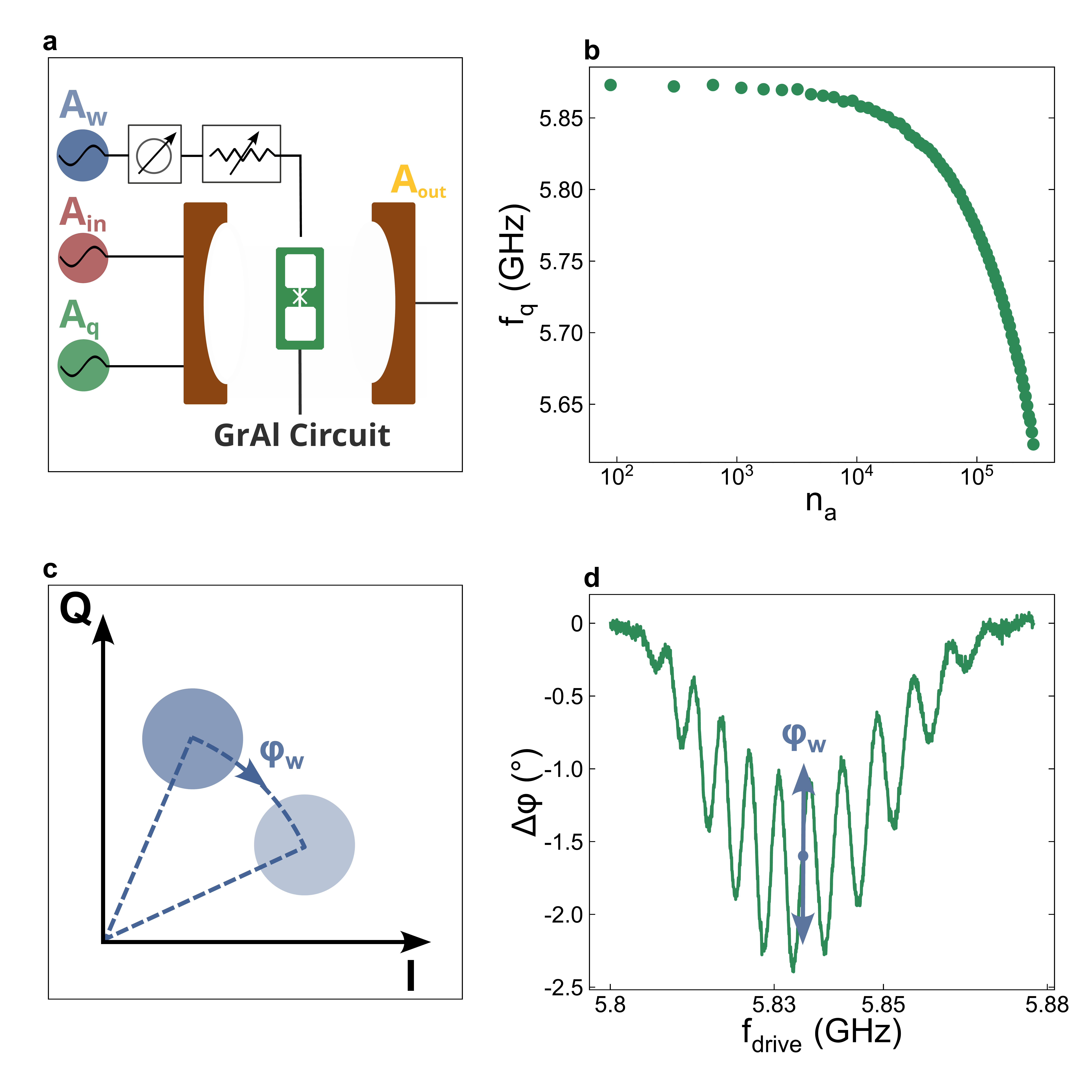}}
\caption{\textbf{Principle of the non-linear interferometer}\newline
a. Block diagram of the non-linear interferometer setup with a cavity and a granular aluminium quantum circuit. The system is driven by 3 different microwave sources driving the circuit, the cavity and simulating the weak microwave signal to be detected. b. Measurement of the frequency dependence of the grAl circuit frequency as a function of cavity number of photons. c. Schematics of the effect of the weak microwave signal on the cavity field in the phase space representation for protocol B. d. Measured Ramsey fringes for the quantum circuit. The Ramsey sequence consists of two $\pi/2$ pulses separated by a free evolution period, during which the quantum phase accumulates. The final state is read out via the coupled microwave cavity. $t_{\pi/2}$ and $t_{ramsey}$ were optimized to 25 ns and 150 ns to optimise the signal to noise ratio. The value of $t_{\pi/2}$ is a bit smaller than the theoretical one, due to the small anharmonicity of the signal.}%
\label{fig:principle}
\end{figure}

\section{EXPERIMENTAL IMPLEMENTATION}\

\subsection{Experimental Setup}

The block diagram of the setup is shown in figure 1a. It comprises a microwave 3-dimensional cavity with a granular aluminum (grAl) circuit mounted in a "transmon" geometry \cite{Thery:24,grAl:23}. There are 3 coherent signals feeding the cavity input: the cavity signal $A_{in}$, resonant with the cavity mode at $6.193$~GHz, the circuit mode $A_q$ driving the grAl circuit in the time domain and the weak signal $A_{w}$. The signal $A_{w}$ has the same frequency as the cavity mode, but phase shifted and strongly attenuated as shown in figure 1a. Figure 1b shows the main resource used in this work, the dependence of the frequency of the quantum circuit as a function of the number of photons in the cavity mode provided by the amplitude $A_{in}$ (see Supplementary for calibration procedures). The frequency has the standard linear dependence for small $n_a$ as expected from the dispersive coupling of the cavity and the circuit \cite{Thery:24}. Such a parametric coupling of the circuit and the cavity enables a phase resolved measurement of the number of photons provided by the amplitude of the weak signal $A_w$ \cite{CottetKontos:24} which "turns" the state of the cavity field in the phase space (In phase (I)-Quadrature phase(Q) plane), as depicted in figure 1c. 

\subsection{Ramsey Interferometry}

This parametric coupling is accurately mapped in our case through Ramsey interferometry of the circuit. The circuit is prepared in a coherent superposition of its states by means of two successive microwave pulses separated by a variable delay. The resulting interference pattern, known as Ramsey fringes, arises from the phase accumulated by the circuit during the free evolution time. A typical Ramsey fringes pattern is shown in figure 1d. 

The interference between the weak probe $A_{w}$ and the cavity drive $A_{\mathrm{in}}$ produces a horizontal shift of these fringes, depending on their relative phase. Consequently, the circuit readout phase of the output signal $A_{\mathrm{out}}$ oscillates accordingly. The Ramsey pulse sequence is detailed in the Appendix B. The timing and amplitudes of the two pulses are optimized to maximize the signal-to-noise ratio.

\subsection{Measurement Schemes}
The main objective is to detect the smallest possible value for $n_w$. To do so, two measurement schemes are considered for detecting this weak signal photon number. Both protocols use a circuit drive at a single frequency, chosen at the point of maximum slope in the Ramsey fringe pattern shown in Figure 1d.
\begin{itemize}
\item \textbf{Protocol A} measures the cavity field directly by recording the in-phase (I) and quadrature (Q) components of the High-Electron-Mobility-Transistor (HEMT) amplifier output. This provides a direct representation of the cavity field trajectory in phase space. In this protocol, the granular aluminium circuit is expected to "preamplify" the signal $n_w$ owing to its non-linearity.

\item \textbf{Protocol B} uses only the dispersive coupling between the granular aluminum circuit and the cavity mode. Here, the phase of the cavity field, given by $\varphi_{out}=\arctan \mathrm{Q/I}$,  carries the information about $n_w$. This protocol a priori works outside the bandwidth of the High-Electron-Mobility-Transistor (HEMT) and is the main amplification resource studied in our work.
\end{itemize}

The two protocols are essentially variations of the same scheme. The main advantage of the latter is that, provided the dispersive shift is sufficiently large, detection can be performed over a frequency range far exceeding the bandwidth of the HEMT amplifier. This aspect is discussed at the end of the main text.

\begin{figure}[tph]
{\small \centering\includegraphics[width=0.65\linewidth,angle=0]{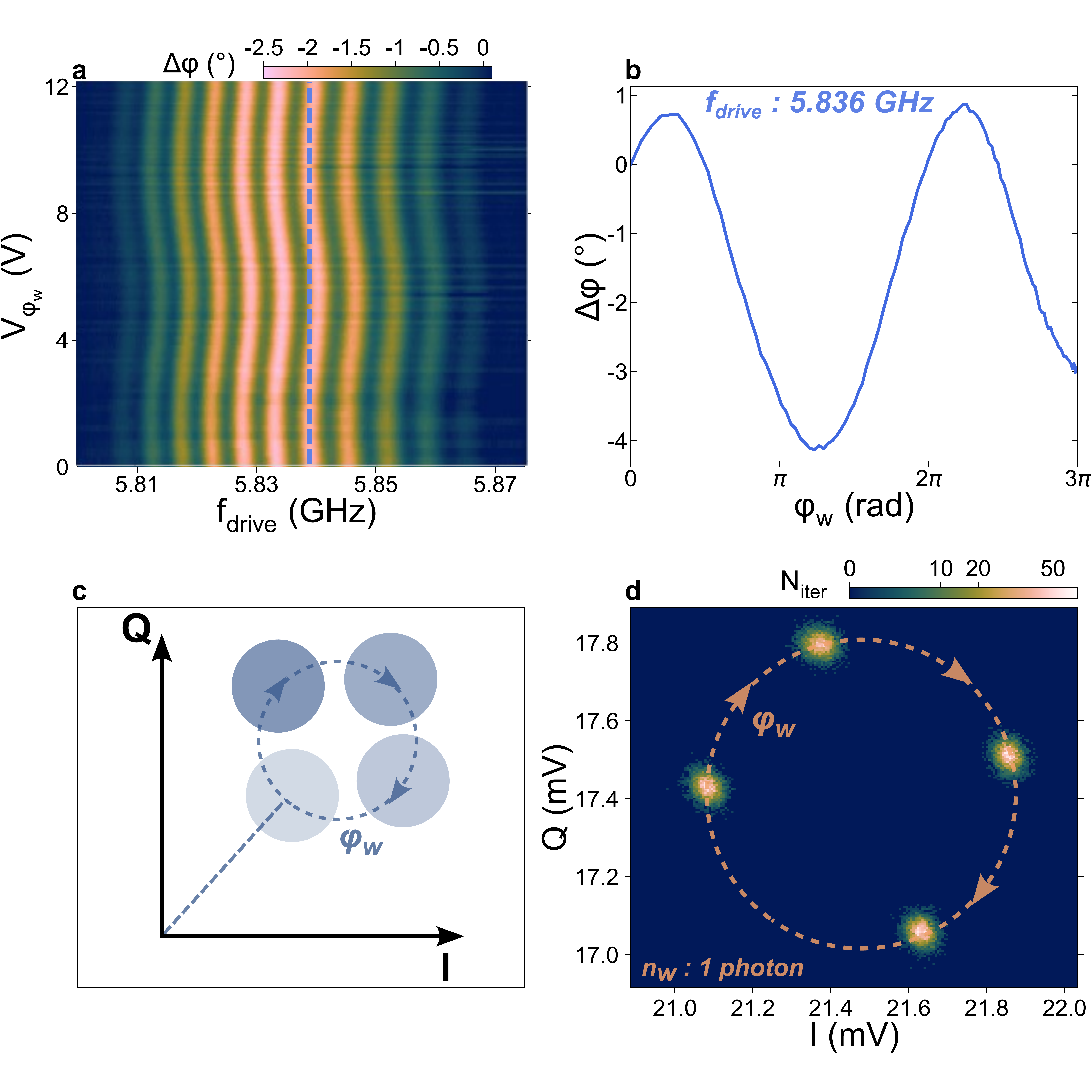}}
\caption{\textbf{Detection of a weak coherent field}\newline
a. Evolution of the Ramsey fringes for $n_w=0.6$ photons as a function of the phase difference of the weak signal and the cavity field. The phase contrast outside the fringes has been subtracted. b. Phase signal corresponding to the vertical cut of signal for $f_{drive}=5.836  GHz$. In this case, the full signal is represented. This explains the differences in oscillation amplitude between panel (b) and panel (a). c. Schematics of the effect of the weak microwave signal on the cavity field in the phase space representation for protocol A. d. Measurement of the evolution of the state of the cavity field as the phase of the weak signal spans $2\pi$ for $n_w=1$ photon.}%
\label{fig:interference}
\end{figure}

\section{DETECTION PRINCIPLE OF THE NON-LINEAR INTERFEROMETER}

The principle of our non-linear interferometer can be demonstrated by measuring the evolution of the Ramsey fringes as a function of the phase $\varphi_w$ of the weak signal $A_w$. The frequency of the circuit is expected to shift as 
\begin{equation} 
\Delta f_q = K_{aq} |\sqrt{n_a}+\sqrt{n_w} e^{i \varphi_w}|^2
\label{eq:deltafq}
\end{equation} 
where we have converted $A_{in}$ and $A_w$ into photon number $n_a$ and $n_w$ respectively, $K_{aq}$ being the cross-Kerr parameter stemming from the self-Kerr $K_q$ of our circuit, estimated to $200\pm50$ kHz from measurement of the Rabi oscillations\cite{Thery:24}. 
The self-Kerr $K_q$ of our circuit can be compared to the theoretical expression $K_0 \approx C \pi a \omega^2 / (j_c V_{\mathrm{gr,Al}})$ proposed by Maleeva et al.\cite{Maleeva:18}, with $C = 3/16$, $a$ being the Al grain size. Using $a = 5~\text{nm}$, we estimate  $K_0\approx 250~\text{kHz}$, which is in good agreement with previous analyses of granular aluminum resonators, where the Kerr nonlinearity is linked to the microscopic granular structure of the film\cite{Maleeva:18}. 

In equation (\ref{eq:deltafq}), there are the two conventional AC-Stark shift terms proportional to $n_a$ and $n_w$, which are used for example in single photon counting experiments \cite{Dixit:19,Braggio:24,Belambois:24} but there is also an interference term modulating as $\varphi_w$. This effect is directly observed in figure 2a, where $\varphi_w$ is changed with the analog phase shifters from $0$ to $2 \pi$ (see calibration in Supplementary). The frequency center of the Ramsey fringes pattern is shown to oscillate as a function of $\varphi_w$. The resulting phase oscillation is shown in figure 2b. 

Protocol A is illustrated in figure 2c:  the evolution of the state of the cavity in the I-Q plane is measured as one sweeps the phase $\varphi_w$ , thanks to the phase shifter. Experimentally, we map directly the field by making histograms in color scale as a function of the I and Q components of $A_{out}$, as depicted in figure 2d. We observe well-defined spots that revolve as $\varphi_w$ spans $2 \pi$. The radius of the circle described is directly proportional to $\sqrt{n_w}$. In figure 2d, each point of the four spots corresponds to a repetition of $N=5000$. The size of these spots is consistent with the noise temperature of our High-Electron-Mobility-Transistor (HEMT) amplifier noise which corresponds to an addition of about $10$ photons in each measurement (see calibration of the noise of the HEMT in the Supplementary).

\begin{figure}[tph]
{\small \centering\includegraphics[width=0.55\linewidth,angle=0]{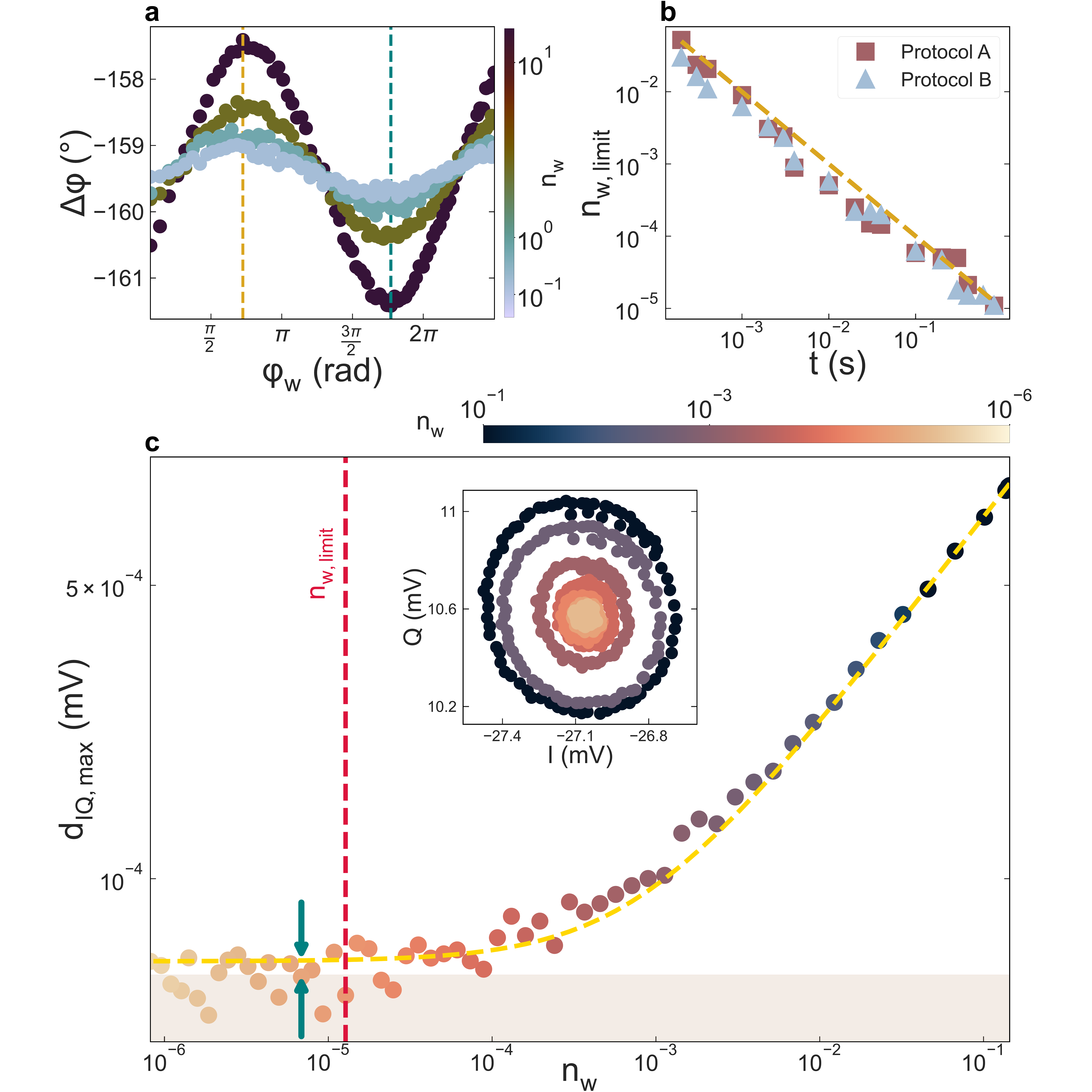}}
\caption{\textbf{Figure of merit of photon measurement}\newline
a. Evolution of the contrast of the Ramsey signal for different values of $n_w$. b. Scaling of the variance of the signal as a function of the measurement time for protocols A and B c. Evolution of the signal in the I-Q plane for  protocol A down to $n_w\approx10^{-6}$ photons in the weak signal, showing a sensitivity of the setup down to several $10^{-5}$. Inset : "Phase circles" in the I-Q plane. The slight elliptical shape is attributed to small phase drifts in the experimental setup. }%
\label{fig:figureofmerit}
\end{figure}

\section{DETECTION FIGURE OF MERIT}

We now turn to the determination of the figure of merit of the detector. We will use both protocols to confirm that they are consistent with each other. The coherent signal is first applied at the readout cavity frequency, where it interferes with the cavity drive injected before the readout, as shown in the Supplementary Information (Section C, left panel of Fig.~S8).

Figure 3a illustrates how the phase modulations evolve as $n_w$ is lowered. Following protocol B, the amplitude of the oscillations decreases accordingly, but we clearly see oscillations down to $n_w=10^{-1}$. We determine the figure of merit of the detector through the evolution of the difference between the minimum and the maximum of the phase contrast as a function of $\varphi_w$. For $n_w$'s which can be detected, we determine the gain of the detector. For $n_w$'s below the detection threshold, we measure essentially the variance $\delta \phi$ of the detection phase. Our detection scheme B is expected to display a linear dependence of the detector signal as a function of $\sqrt{\varphi_w}$ (constant gain). A practical way to determine unambiguously the detection threshold is therefore to find the intersection between the average $\overline{\delta \phi}$ and the line of constant gain. This is done in the supplementary. 

In protocol A, the detector signal is simply the diameter $d_{IQ}$ of the "phase circles"  in the I-Q plane above the detecting threshold which becomes the size of the coherent state spot of the cavity field, directly related to the variance of the two quadratures. The detecting threshold is defined as the value of $n_w$ when $d_{IQ}$ exceeds by $5\%$ the variance baseline. Interestingly, one can already study the dynamics of the measurement as a function of the measurement time which is taken as the number of pulse sequences multiplied by the sequence length which is $4 \mu s$ throughout our study. 

This is shown in figure 3b for protocol A and B. One observes that both protocols agree quantitatively as far as the variance is concerned. As expected for independent measurements, the variance, represented in figure 3b, scales like $1/t$ where $t$ is the measurement time corresponding to the pulse sequences. In the quantum metrological sense, we do not reach the Heisenberg limit but we are well below the $n_{w}=1$ threshold. Our measurements show in particular that we can detect below $n_{w}=1$ already for a measurement time of $200 \mu s$. This is one of the main results of our work. The figure of merit of our detector is the threshold detection for a given measurement time which we show in figure 3c, for $400$~ms. The detection signal $d_{IQ}$ decreases linearly as a function of $\sqrt{n_w}$ as expected and levels off at the variance of the quadratures. The small distortion observed in the inset corresponds to a genuine phase shift, caused by fluctuations in the laboratory conditions. The dashed line is the expected function $f(n_{w}) = A \sqrt{n_{w} + \frac{(\delta n_{w})^2}{t}}$ with $ A \sim 2 \times 10^{-3}$, $\delta n_{w} \sim 10^{-1}$. We can detect a signal as low as $n_{w}\approx10^{-5}$. From the intercept of the square root scaling in figure 3b, we can determine that this corresponds to a figure of merit of $1.2\times 10^{-22} W/\sqrt{Hz}$. This is similar to what has been recently found for single photon detectors \cite{Belambois:24} and about two orders of magnitude better than single photon detection schemes used recently in dark matter quantum sensing setups\cite{Dixit:19}. Interestingly, both Single Microwave Photon Detector (SMPD) setups were equipped with traveling wave parametric amplifiers (TWPAs), whereas our setup is only equipped with a High Electron Mobility Transistor (HEMT) amplifier which has at least 10 times larger noise temperature. This shows further the interest of the phase resolved method. Note however that the grAl circuit has only a modest effect as shown in figure 10 of the Appendix C. There is a preamplification of the signal before the HEMT thanks to the Ramsey fringes of about $10\%$ which reaches an optimum, arising from the visibility of the Ramsey fringes. Such a modest enhancement shows the role of the grAl in protocol A. It certainly could be enhanced in a more traditional parametric amplifier setup in order to get more gain but this is outside the scope of the present study. Overall, similar to the amplification schemes with SMPDs or TWPAs, the Josephson non-linearity is at work as a preamplification scheme. In our case, although our setup has the bandwidth of our HEMT, the dispersive detection enables to detect microwave signals of very different frequencies, thus enabling potentially very wide band scanning range (larger than $10 GHz$) \cite{CottetKontos:24}. Finally, our use of a grAl circuit, which is 'photon resilient' owing to its weak anharmonicity, enables a large dynamic range of more than 30 dB as one can read-off from figure 3c and 4a. 

\begin{figure}[tph]
{\small \centering\includegraphics[width=0.65\linewidth,angle=0]{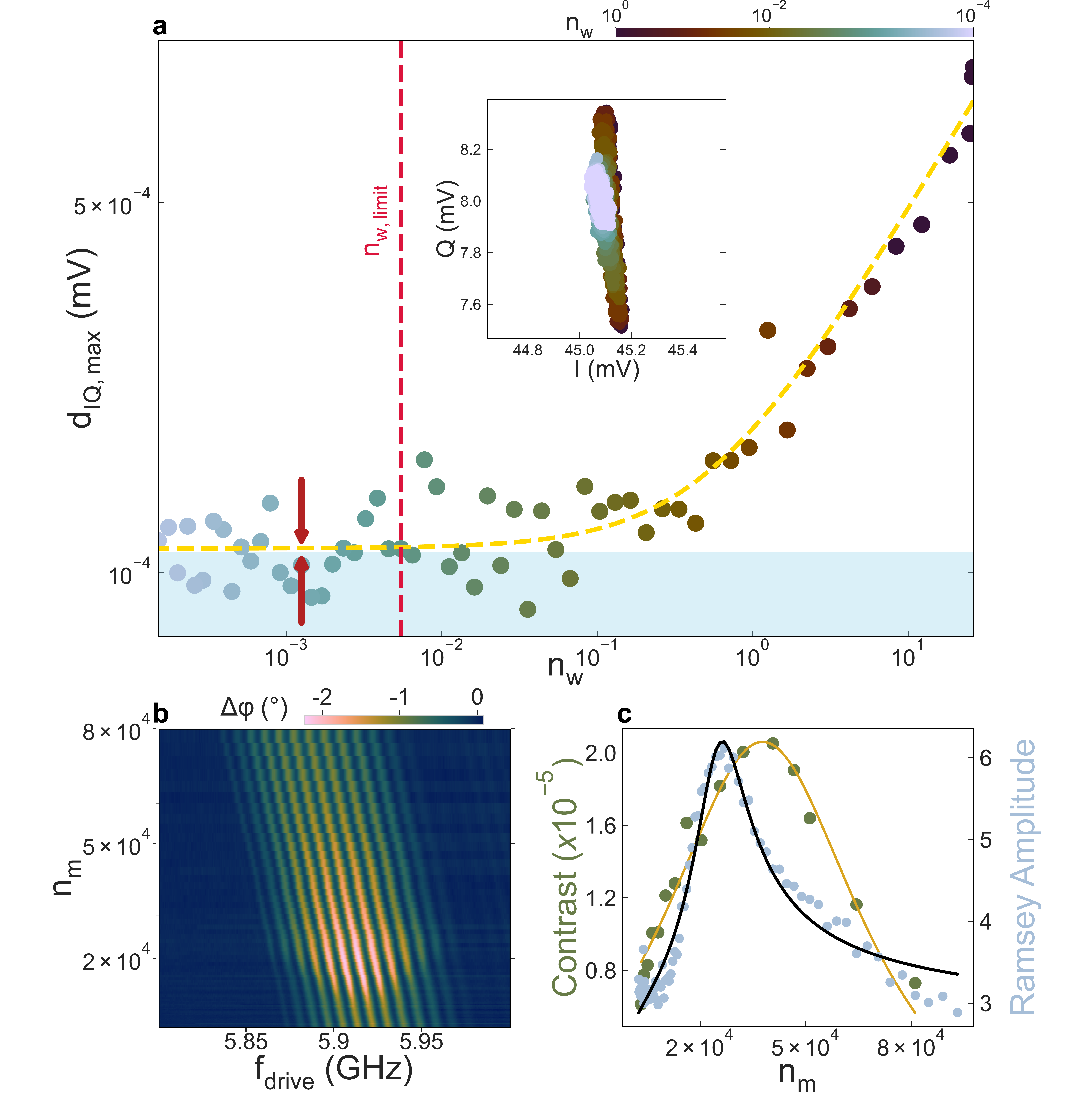}}
\caption{\textbf{Evading  the shot noise limit}\newline
a. Evolution of the contrast of the Ramsey signal for different values of $n_w$ for protocol B down to $n_w\approx10^{-4}$. For this non-resonant detection scheme, the trajectory of the state of the field in the I-Q plane is consistent with a piece of circle. The experiment is done at $f_{ramsey} = 5.919$GHz. b. Evolution of the Ramsey fringes as a function of the pump power c. Evolution of the contrast as a function of the pump power. The Ramsey amplitude is fit with a lorentzian whereas the interferometer contrast is fit with a nonlinear function given the maintext.}%
\label{fig:optimum}
\end{figure}

\section{EVADING THE SHOT NOISE LIMIT}

The nonlinear-interferometer principle stems from the dispersive interaction between the quantum circuit in grAl and the use of a pump filling a cavity with $n_a$ photons. This can be used to probe any mode dispersively coupled  to the circuit in principle so long as there is a sizable dispersive shift. 

An auxiliary cavity mode at $5.065~\mathrm{GHz}$, distinct from the readout mode, is now excited, with the coherent signal set to this frequency. Protocol A is then repeated for this mode, and the corresponding measurement is shown in Fig.~4. The pulse sequence used for this measurement differs from the previous one (see Supplementary Information, Fig.~S8, right): a continuous drive excites the lower mode, while the readout mode is used solely for the circuit readout. This configuration accounts for the frequency shift observed in the Ramsey pattern compared to the earlier measurements.

The result of protocol A is presented in figure 4a, with the new circuit frequency set to $5.919$ GHz. The resulting trajectory in the I-Q plane, shown in inset of figure 4a, is consistent with the cartoon of figure 1c. Interestingly, we also observe a signature of the $\sqrt{n_w}$ scaling in the threshold for detection (shown in the supplementary). In this case, the sensitivity appears to be more around $n_w \lesssim 10^{-2}$. The effect of the pump power on the Ramsey fringes is shown in figure 4b. Apart from the shift of the Ramsey pattern towards low frequency, we observe a maximum of the interference contrast. Consistently, the resulting gain of  interferometer peaks at about $P_m\approx 0 dBm$. It is interesting to note that the maximum coincides with the maximum of contrast of the Ramsey fringes. Such an optimum in the power of the interferometer is a phenomenon alike measurement induced dephasing predicted for anharmonic oscillators \cite{Cottet:20,CottetKontos:24}. Interestingly, one can fit the contrast with a simple empirical function of the form $A+B\frac{n_m^E}{\sqrt{((n_m - n_{m0})^2 + C^2)}}$ with $A \approx 2.54$, $B \approx 115$, $E \approx 0.54$, $C \approx 7540$ and $n_{m0} \approx 2.5\times 10^4$. There is an optimum for $n_m$ (around  $n_{m0} \approx 2.5\times 10^4$) for the detector contrast. Such a measurement contrasts with the expected $\sqrt{n_m}$ scaling one should have for a conventional linear interferometer subject to the shot noise of the reference beam characterized by the number of photons $n_m$. In addition, the shot noise limit usually refers to the number of photons in the read-out of the interferometer, which is kept constant here. Therefore, our non-linear setup evades the shot noise limit in several ways.

\section{DETECTING AT HIGHER MAGNETIC FIELD}

\begin{figure}[tph]
{\small \centering\includegraphics[width=0.65\linewidth,angle=0]{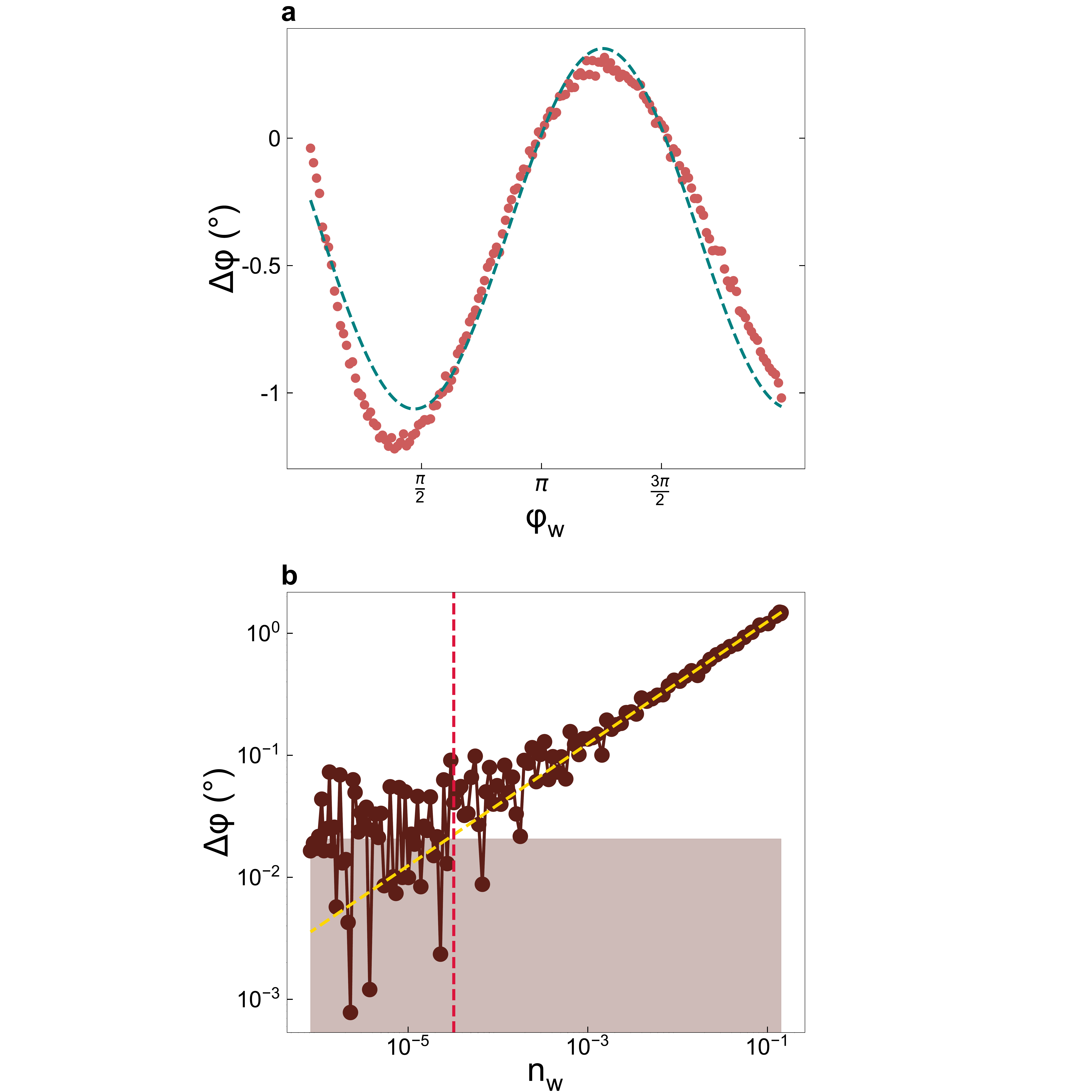}}
\caption{\textbf{Detecting at higher magnetic field}\newline
 a. Evolution of the Ramsey signal contrast for different values of $n_w$ using Protocol B, measured at $n_w \approx 1$ and $B_0 = 100$ mT.
b. Decrease of the Ramsey fringe contrast as $n_w$ is reduced, using Protocol B. A sensitivity limit of $5 \times 10^{-4}$ photon is reached in a time of acquisition of $400 ms$. }\label{fig:fig5}
\end{figure}
Before concluding, it is interesting to showcase the robustness of the interferometer demonstrated here with respect to environmental perturbations. We have already demonstrated that our interferometer can operate with a large photon background, of about  $10^4-10^5$. Another property linked to our use of a granular aluminum based quantum circuit is the magnetic field resilience of our setup. We show in figure 5 that we can implement the protocol B under a magnetic field as high as $100 mT$. As granular aluminum quantum circuit are magnetic field resilient \cite{Thery:24} and can now be operated up to more than $1T$ \cite{grAl:25} or used for magnetic field resilient parametric amplification\cite{gralAmp:24}, we could imagine to operate our setup at even higher magnetic fields. This would be particularly relevant for axion quantum sensing applications which require a large magnetic field to convert potential axion signals into cavity photons.

\section{CONCLUSION}

In conclusion, we have demonstrated a protocol for measuring weak microwave signals using a non-linear quantum interferometer. One immediate application of our findings is sensing microwave fields in haloscopes for axion dark matter. It is interesting to note that our use of granular aluminum which have been shown to be strongly magnetic field resilient lift the constraint of some setups using superconducting quantum bits \cite{Dixit:19}. Our method of non linear interferometry which bares strong analogies with recently proposed theoretical protocols \cite{Kerrmedium:22,CottetKontos:24} may be transposed to interacting fermions, for example in Majorana\cite{Majorana25} or Fractional Quantum Hall interferometers\cite{Anyon20}, with applications to anyon detection and braiding, where electron-electron interactions would play the role of the Kerr photon-photon interaction.

\clearpage
\begin{acknowledgements}
 We thank I. Irastorza, K. Petraki, Z. Leghtas, R.T. D'Agnolo and P. Fayet for fruitful discussions. We also gratefully acknowledge the RADES collaboration. This work is supported by the QRADES Quantera project and the DarkQuantum ERC project.
 \end{acknowledgements}


\section*{Appendix A: Description of the superconducting circuit}\label{section:calc_details}
\subsection{Nanofabrication of the Granular Aluminium Josephson Junction}

The granular aluminium junction (GrAl) is fabricated in a single nanolithography step, using two evaporation angles. The sample fabrication and preliminary characterization have already been reported in \cite{Thery:24}.

\subsection{Experimental Setup \& Detection Protocol}

\begin{figure}[htbp]
    \centering
    \includegraphics[width=1.0\textwidth]{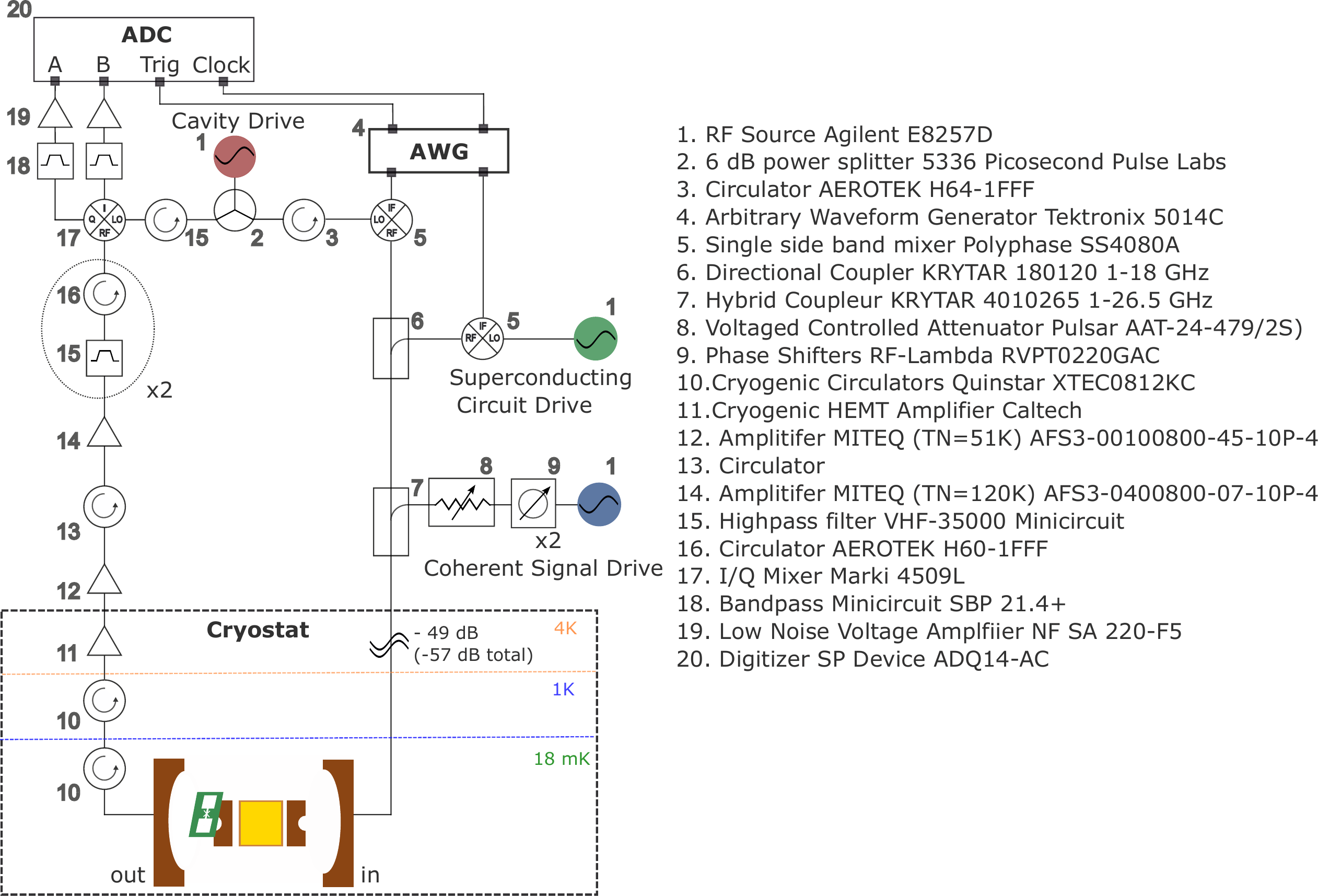}
    \caption{Detailed Experimental Setup (top)}
    \label{fig:exp}
\end{figure}

The detection setup is mounted in a dilution refrigerator, operating at a base temperature of 18 mK. The cryostat wiring is detailed in reference \cite{bruhat2016}.  The detailed experimental setup is illustrated in figure \ref{fig:exp} : which shows the wiring scheme at each temperature stage of the cryostat, where RF signals are attenuated and thermalized. 

We perform heterodyne detection on the cavity and superconducting circuit with the following RF signals : the local oscillators (LO) are generated by two different RF sources operating at a frequency $\omega_{LO}$. An arbitrary waveform generator generates two different intermediate frequencies (IF) at a frequency $\omega_{IF}$, respectively $20$ MHz and $50$ MHz. These signals are then processed by two single-sideband (SSB) mixers, which upconvert them to $\omega_{LO} + \omega_{IF}$ before sending them into the cryostat. 

To test the detection setup, a perturbative coherent signal is introduced using a continuous tone generated by a third RF source.

\newpage

\begin{figure}[htbp]
    \centering
    \includegraphics[width=0.45\textwidth]{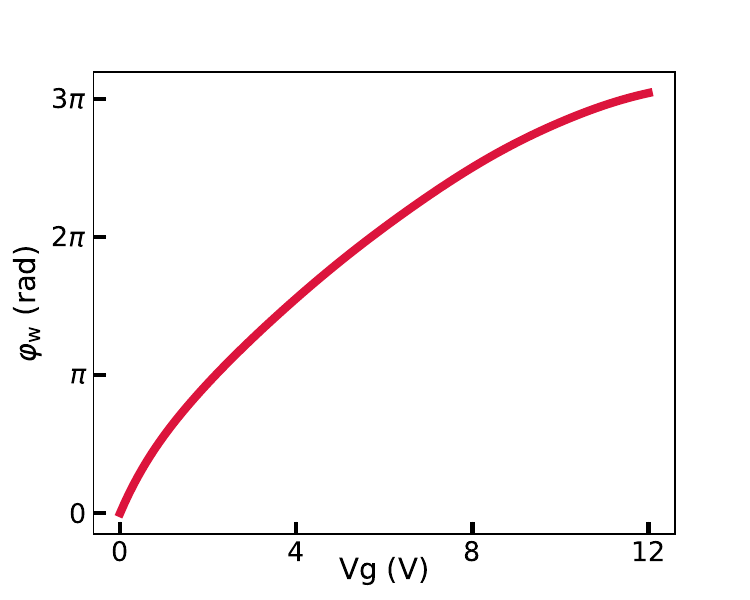}
    \includegraphics[width=0.45\textwidth]{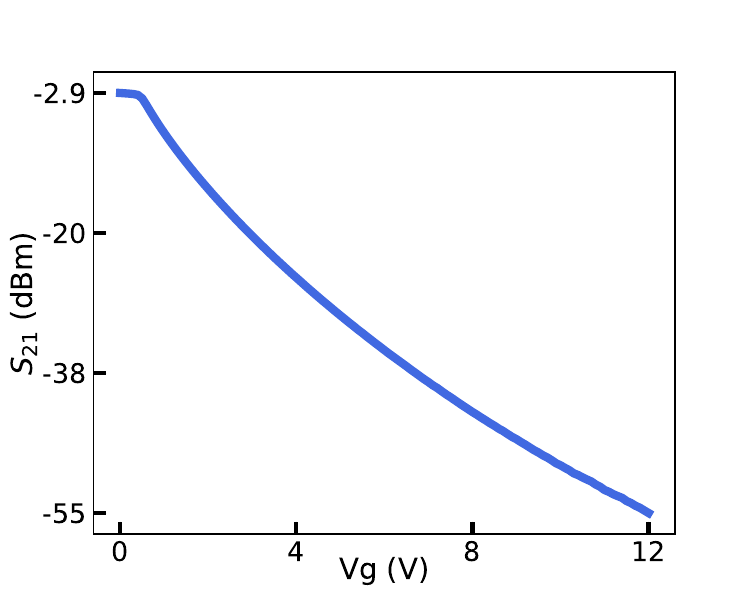}
    \caption{Calibration of the Perturbative Coherent Signal: (Left) Calibration of the two phase shifters as a function of the applied voltage. The plot shows the relationship between the control voltage (ranging from $0$ V to $12$ V) and the resulting phase shift, covering a full phase range from $0$ to $3 \pi$. (Right) Calibration of the variable attenuator as a function of the applied voltage. This plot displays the relationship between the control voltage (ranging from 0 V to 12 V) and the obtained attenuation, which will be added to the fixed attenuation (93 dB).}    \label{fig:cal}
\end{figure}

Each DC source (Yokogawa GS200) controls the two phase shifters, to simulate the oscillating phase of the perturbative coherent signal. A voltage ranging from 0 V to 12 V corresponds to an oscillating phase varying from $0$ to $3\pi$, with the calibration shown in the left figure in Figure \ref{fig:cal}. Each phase shifter introduces an insertion loss of 5 dB. Finally, the signal is attenuated incrementally using a voltage-controlled variable attenuator in combination with a fixed attenuation of 69 dB. The attenuation calibration is presented in the right figure in Figure \ref{fig:cal}.

\newpage

\subsection{Pulse Sequences}

\begin{figure}[htbp]
    \centering
    \includegraphics[width=0.45\textwidth]{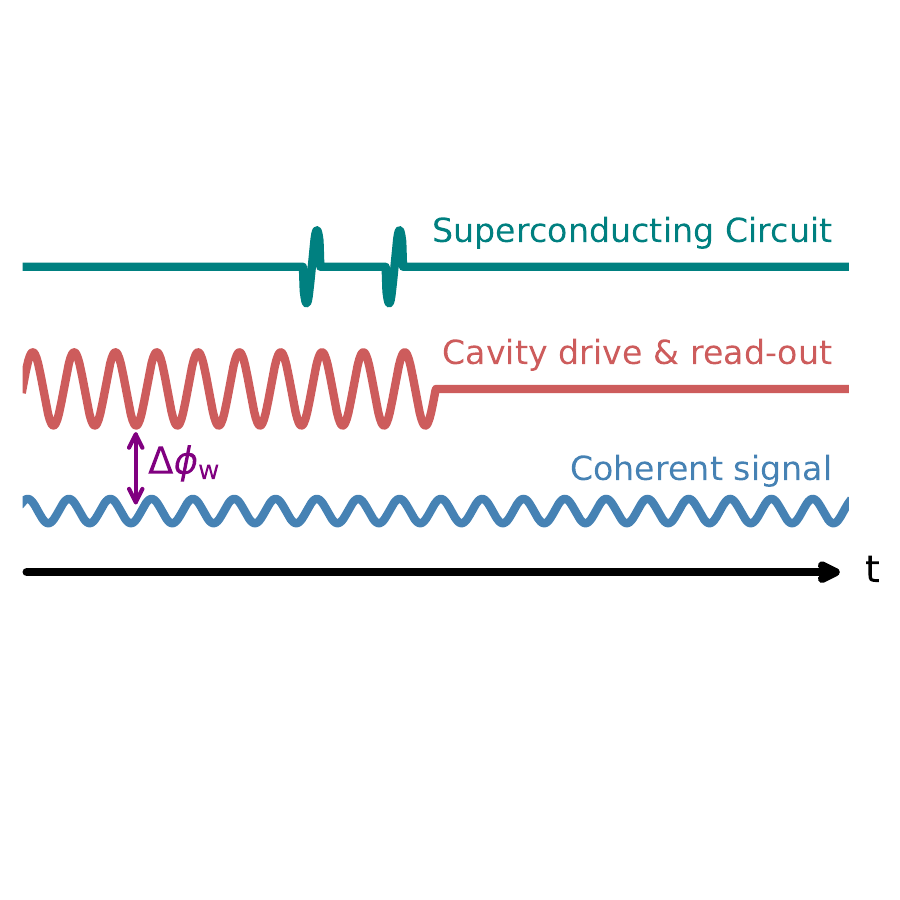}
    \includegraphics[width=0.45\textwidth]{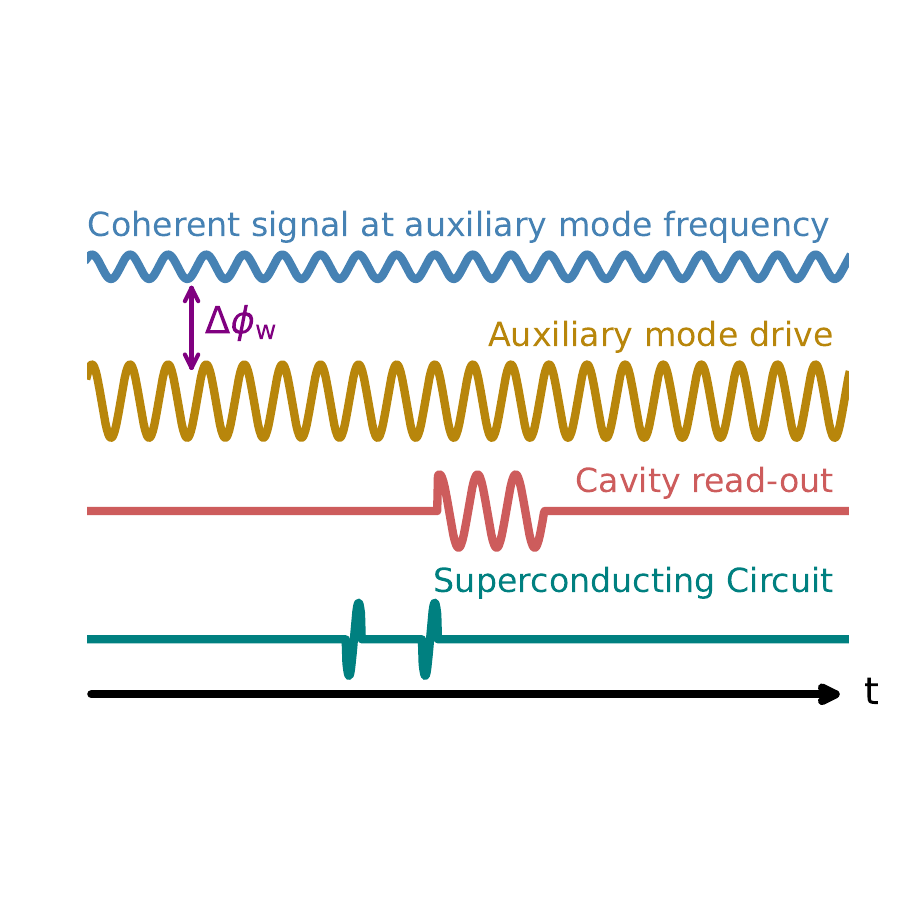}
    \caption{Pulse sequences used in the experiment. (Left) Standard configuration with a single cavity mode for both control and readout. (Right) Modified setup using an additional cavity mode to decouple the coherent drive from the readout path, as used in Fig.~4 of the main text.}    \label{fig:pulse_seq}
\end{figure}

A schematic representation of the pulse sequences used throughout this work is shown in Fig.~\ref{fig:pulse_seq}.

To optimize the detection process, the coherent signal is first generated at the cavity resonance frequency, where the transmission is maximized. A drive pulse is applied at this frequency using an RF source, and its amplitude is modulated at 20 MHz via an Arbitrary Waveform Generator (AWG). A second RF source is employed to drive the superconducting circuit, with its output modulated at 50 MHz by the AWG to enable the implementation of a Ramsey sequence. Additionally, a coherent drive at the cavity frequency is continuously applied using a third RF source. This signal is attenuated by the variable attenuator, and its phase is adjusted using two phase shifters. The interference relevant to the measurement occurs between the cavity drive and the coherent signal.

The detection protocol consists in a Ramsey pulse sequence with a total duration of $t_{\mathrm{sequence}} = 4 \mu\mathrm{s}$. In the initial stage, the cavity is populated with photons at its resonance frequency during a pulse of duration $t_{\mathrm{pulse}} = 2 \mu\mathrm{s}$. At $t=t_{pulse} - t_{ramsey} - 2t_{\frac{\pi}{2}}$, a first $\frac{\pi}{2}$-pulse is applied to the superconducting circuit for a duration $t_{\frac{\pi}{2}}$, placing the circuit into a superposition of two distinct states.  The first $\frac{\pi}{2}$-pulse is followed by a free evolution during $t_{ramsey}$, and another $\frac{\pi}{2}$-pulse for measurement. The system is then measured via the cavity during $t=t_{meas}=0.3\mu s$, and an additional post-processing delay ($1 \mu s$) ensures the circuit returns to its initial state, completing the sequence at  $t_{sequence}$. Both timing and amplitude parameters were optimized for this measurement, and the circuit photon number, $n_q$ remains fixed at $n_q = 3 \times 10^4$ photon. Typically, $t_{ramsey} = 150$ns and  $t_{pulse} = 25$ns. 

Simultaneously, a continuous tone from an RF source is applied to simulate a perturbative coherent signal, $\varphi_{\mathrm{w}}$. Its phase varies from $0$ to $2\pi$ using two phase shifters, resulting in the circular pattern in the I-Q plan.

To improve this setup, a fourth RF source is introduced to excite an additional cavity mode. In this configuration, the control of the superconducting circuit and the readout of the cavity are completely decoupled from the coherent drive. A high-power continuous-wave drive is applied at a distinct cavity frequency (typically 5.065 GHz) to excite a second cavity mode. Detection is then performed on this auxiliary mode, with a fourth RF source generating a coherent signal that is again attenuated and phase-controlled using the same active components described above. This modified pulse sequence corresponds to the one used in Fig.~4 of the main text.

In the first scenario, the resulting signal forms a circular trajectory in the IQ plane under phase shifts, with the overall sensitivity determined by both the linear amplifier and the circuit addition. In contrast, when the coherent signal is injected into a different cavity mode, the phase shift manifests as a smaller, partial rotation in the IQ plane, and the sensitivity is then solely attributed to the circuit.

\newpage

\subsection{Photon number determination}

We detail the derivation we used to estimate the number of photons inside the cavity from the power sent inside it.

To estimate the number of photons in the setup, several conversions are required. Experimentally, $V_{pp}$ signals are applied to the cavity and superconducting circuit RF drives, while the perturbative coherent signal is a continuous tone from a RF source. The power levels are converted to $dBm$, then $Watts$.\\

\begin{equation} \label{eq_ni}
\begin{split}
    P_{in,a(dBm)} &=  10 \times \text{log}_{10}(A_{c(Vpp)}^2 \times 2.5)  - 10 - 3 - 3 - 6 \\
    P_{in,b(dBm)} &=  10 \times \text{log}_{10}(A_{q(Vpp)}^2 \times 2.5)  - 20 - 3 \\
    P_{in,w(dBm)} &=  A_{i,var}  - A_{i,fixed} - 20 - 3 - 3 - 10  \\
    P_{in,m(dBm)} &=  P_m  - 20 - 3 - 3 - 10  
    \end{split}
\end{equation}

The cavity line includes a fixed attenuation of $10 dB$, passes through a single-sideband modulator (SSB) with an insertion loss of $6 dB$, and two hybrid couplers with an insertion loss of $3 dB $ each.

The superconducting circuit line has a fixed attenuation of  $20 \text{ }dB $ and also passes through the last hybrid coupler.

The perturbative coherent signal line is generated at $0 \text{ }dBm $. It then goes through a variable attenuator $A_{i,var}$, a fixed attenuation of $A_{i,fixed}$, a first directional coupler with an insertion loss of $23 \text{ }dB $, and finally a hybrid coupler with an insertion loss of $3 \text{ }dB $.

The implementation of the fourth RF source, described in the final section of the paper, includes a hybrid coupler with an insertion loss of  $3 \text{ }dB $. 

We can then convert the power values in $W$ : 
\begin{equation}
\begin{split}
    P_{in,i(W)} &=  10^{(P_{in,i(dBm)}-30)/10} \\
     \end{split}
\end{equation}

We now consider the GrAl superconducting circuit of angular frequency $\omega_0$, coupled to a cavity mode at the frequency $\omega_r$. With the same approximation as in \cite{Thery:24}, we get following Hamiltonian: 

\begin{equation} \label{eq_ni}
\begin{split}
   H &= \hbar \omega_{r} \hat{a}^{\dagger} \hat{a} + \hbar \omega_0 \hat{b}^{\dagger} \hat{b} + \hbar g (\hat{a}^{\dagger } \hat{b} + \hat{a} \hat{b}^{\dagger}) - \hbar \frac{K}{2} \hat{b}^{\dagger} \hat{b}^{\dagger} \hat{b} \hat{b}
     \end{split}
\end{equation}

We derive the equation of motion of $\hat{a}$ and $\hat{b}$ using the input-output formalism. We consider that the cavity is coupled to the input and output lines via two ports of equal coupling constants $\kappa_p$. A drive signal is sent through the input port at the frequency $\omega_{RF}$, and is described as a source term $\hat{b}_{in}(t)$. The associated power is then defined as $P_{in}^{(W)} = \hbar \omega_{RF} \langle b_{in}^{\dagger}(t) b_{in}(t) \rangle$. \\

The total loss decay inside the cavity is defined as : $\kappa = 2 \kappa_p + \kappa_0$, with $\kappa_0 $ the internal losses inside the cavity. The derived equations of motion are then :

\begin{equation} \label{eq_da}
\begin{split}
   \frac{d }{dt} \hat{a}(t) &= \frac{i}{\hbar} [ \hat{H}, \hat{a} ] - \frac{\kappa}{2} \hat{a} - \sqrt{2 \pi \kappa_p} \hat{b}_{in}(t) \\
    \frac{d }{dt}  \hat{a}(t) &= - i\omega_r \hat{a}(t) - ig \hat{b}(t)  - \frac{\kappa}{2} \hat{a} - \sqrt{2 \pi \kappa_p} \hat{b}_{in}(t) 
     \end{split}
\end{equation}
with $\hat{b}_{in}(t)$ the source term due to the drive at the input port. 

Now if we consider a coherent drive at a frequency $\omega_{RF}$, we can write $\langle \hat{b}_{in}(t)\rangle = \bar{b}_{in} e^{- i\omega_{RF}t} $,  $\langle \hat{a}_{in}(t)\rangle = \bar{a} e^{- i\omega_{RF}t} $, and   $\langle \hat{b}_{in}(t)\rangle = \bar{b} e^{- i\omega_{RF}t} $. The number of photon inside the cavity is defined by equation \ref{eq_da} at $\omega_{RF} = \omega_r$ becomes : 

\begin{equation}
\begin{split}
     \langle a^{\dagger} a \rangle &= 2\pi\frac{\kappa_p}{\left( \frac{\kappa}{2} \right)^2} \frac{P_{in}^{(W)}}{\hbar \omega_{RF}} \\
\end{split}
\end{equation}

The values of $\kappa_{a}$,  $\kappa_{p,a}$, $\omega_{RF}$ are determined from the resonance of the cavity. 

The output power going out of the cavity is defined as :

\begin{equation}
\begin{split}
     P_{out}^{{(W)}} &= \hbar \omega_0 \kappa_p   \langle a^{\dagger} a \rangle = \frac{4 \kappa_p^2}{\kappa^2} P_{in}^{(W)} \\
\end{split}
\end{equation}

We find the relation : 

\begin{equation}
\begin{split}
     \kappa_p &= \frac{\kappa_a}{2} \sqrt{\frac{P_{out}^{(W)}}{P_{in}^{(W)}}} \\
\end{split}
\end{equation}

For $P_{out}^{(W)}$ and $P_{in}^{(W)}$, we have the following formula : 
\begin{equation}
\begin{split}
     P_{out}^{(meas)} &= G_0 P_{out}^{(W)} \\
     P_{in}^{(W)} &= 10^{-A_i/10} P_{in,RF}^{(W)} \\
\end{split}
\end{equation}


Defining $S_m$ the measured transmission of the cavity (in dB), $A_i$ the total attenuation of the input lines in the cryostat (57 dB), $G_0$ the amplification of the output line (62 dB),  $P_{RF}^{(dBm)}$ the measurement power (in dBm) delivered by the RF setup, we get

\begin{equation}
\begin{split}
   10 \log{\frac{P_{out}^{(W)}}{P_{in}^{(W)}}} &= S_m^{(dB)} - 10 \log(G_0^{(W)}) - 10 \log(P_{in,RF}^{(W)}) + A_i^{(dB)}  \\
\end{split}
\end{equation}

In addition, for $\kappa_p$, we have : 
\begin{equation}
\begin{split}
     \kappa_p &= \frac{\kappa_a}{2}  \times 10^{(S_m - G_0 - P_{in,RF}^{(dBm)} +A_i - 30)/20} \\
\end{split}
\end{equation}
with $S_m$, $A_i$ and $G_0$ in $dB$ and $P_{in,RF}^{(dBm)}$ calculated with the formula (\ref{eq_ni}).

The parameters extracted for the main cavity mode are:
\begin{equation}
\begin{split}
    \kappa_a &= 2 \pi \times 5.2 \text{ MHz}\\
    \kappa_{p,a} &= 2 \pi \times 1.32 \text{ MHz}\\
    \omega_{RF,a} &= 2 \pi  \times 6.193 \text{ GHz} \\
\end{split}
\end{equation}

Using a vector network analyzer (VNA) calibration, we also determine the parameters corresponding to the additional cavity mode used in Fig.~4 of the main text:

\begin{equation}
\begin{split}
    \kappa_{m} &= 2 \pi \times 12.8 \text{ MHz}\\
    \kappa_{p,m} &= 2 \pi \times 1.37 \text{ MHz}\\
    \omega_{RF,m} &= 2 \pi  \times 5.065 \text{ GHz} \\
\end{split}
\end{equation}

Finally, the photon number is estimated using the previously established formula :

\begin{equation}
    \boxed{
     n_i^0 = \frac{8 \pi   P_{in}^{(W)}\kappa_{p}}{\hbar \omega_{RF} \kappa_a^2}
     }
\end{equation}

\section*{Appendix B: Analytical modeling details for Ramsey Fringes}
\label{section:calc_details}

\subsection{Description of the model}

In this study, we consider a granular aluminium circuit interacting within an interferometric setup. The annihilation operators of the superconducting circuit are denoted by $\hat{\beta}$.

\subsection{Semi-classical derivation}

We derive the equation of motion of $\beta$ in the classical regime : 
\begin{equation}\label{eq:ramsey}
    \begin{split}
    \dot{\beta}(t) &= - i \omega_q \beta(t) - \frac{\Gamma}{2} \beta(t) - \epsilon_{drive}(t) e^{-i \omega_d t} \\
    \end{split}
\end{equation}

with : 
\begin{equation}
    \begin{split}
    \epsilon_{drive}(t) &= \epsilon_{0}(t) + \epsilon_{0}(t+ \tau)\\
    \beta &= \beta_1(t) + \beta_2(t)
    \end{split}
\end{equation}

By applying the Rotating Frame Approximation, we obtain : 

\begin{equation}
    \begin{split}
    \bar{\beta} &= i (\omega_d - \omega_q) \beta - \frac{\Gamma}{2} \beta - \epsilon_{drive} \\
    \end{split}
\end{equation}

We choose the form of $\beta(t)$ as : 

\begin{equation} \label{eq:4}
    \begin{split}
    \beta(t) &= e^{i \delta t } A(t)\\
    \end{split}
\end{equation}
with $\delta = (\omega_d - \omega_q) + i \frac{\Gamma}{2}$

We replace this equation in \ref{eq:ramsey}, which gives : 
\begin{equation}
    \begin{split}
    \dot{A(t)} + A(t) [..] &= A(t) [..] - i \epsilon_{drive} e^{- i \delta t}  \\
    \end{split}
\end{equation}

This equation gives the form of $A(\tau)$ as :
\begin{equation}
    \begin{split}
      A(\tau) &= \int^{\tau}_0 - i \epsilon_{drive}(t) e^{- i \delta t} dt + A_0 \\
    \end{split}
\end{equation}
Now $\epsilon_{drive}(t)$ is considered as a square pulse of amplitude $\epsilon_0$.
This gives : 
\begin{equation}
    \begin{split}
    A(t) =  \frac{-i \epsilon_0}{-i (\omega_d - \omega_q) + \frac{\Gamma}{2}} \left( e^{-\delta  \tau} -1\right) + A_0
    \end{split}
\end{equation}

We assume that $A_0 = 0$. We find the expression of $\beta$ as a function of time: 

\begin{equation} \label{eq:8}
    \begin{split}
    \beta(t) &= \frac{-i \epsilon_0}{-i (\omega_d - \omega_q) + \frac{\Gamma}{2}} \left( e^{- i \delta \tau} -1\right) e^{i \delta t} \\
    \end{split}
\end{equation}

Thus, starting again the same procedure as \ref{eq:4} - \ref{eq:8} right after the end of the Ramsey Sequence ($\tau_R + 2 \tau$), we have : 

\begin{equation}
\begin{split}
    \beta(\tau_R + 2 \tau) &=  \left(\int_{0 }^{ \tau} - \epsilon_0 e^{- i \delta t } dt + e^{i \varphi} \int_{\tau_R + \tau}^{\tau_R + 2 \tau} - \epsilon_0 e^{- i \delta t } dt \right)  e^{i \delta ( \tau_R + 2 \tau)} \\
    &=  \left(\frac{-i \epsilon_0 }{-i \delta} \left( e^{- i \delta \tau}  - 1 \right)  + \frac{-i \epsilon_0 e^{i \varphi}}{-i \delta} \left( e^{- i \delta (\tau_R + 2 \tau)}  - e^{- i \delta (\tau_R + \tau)} \right)\right) e^{i \delta ( \tau_R + 2 \tau)}   \\
    & = \frac{\epsilon_0 }{ \delta} \left(e^{i \delta (\tau + \tau_R)} -   e^{i \delta (\tau_R + 2 \tau)} \right)+ \frac{ \epsilon_0 }{ \delta}  e^{i \varphi} \left(1 - e^{i   \delta  \tau} \right) \\
    & = \frac{ \epsilon_0}{ \delta} (e^{i \varphi } + e^{ i \delta (\tau_R + \tau)}) (1 -  e^{i \delta \tau})  \\
\end{split}
\end{equation}

We find the analytical value of the phase contrast of the circuit, corresponding to the Ramsey fringes : 

\begin{equation} 
    \boxed{
    |\beta|^2 =  \frac{ \epsilon_0^2}{| \delta |^2} \left| 1 - e^{ i \delta \tau} \right|^2 \left| e^{i \varphi } + e^{i \delta (\tau_R + \tau)} \right|^2
    }
\end{equation}

\newpage

\section*{Appendix C: Additional data}

\subsection{Pump optimum : Influence of the number of photons $n_{a}$ in the cavity }

\begin{figure}[htbp]
    \centering
    \includegraphics[width=0.6\textwidth]{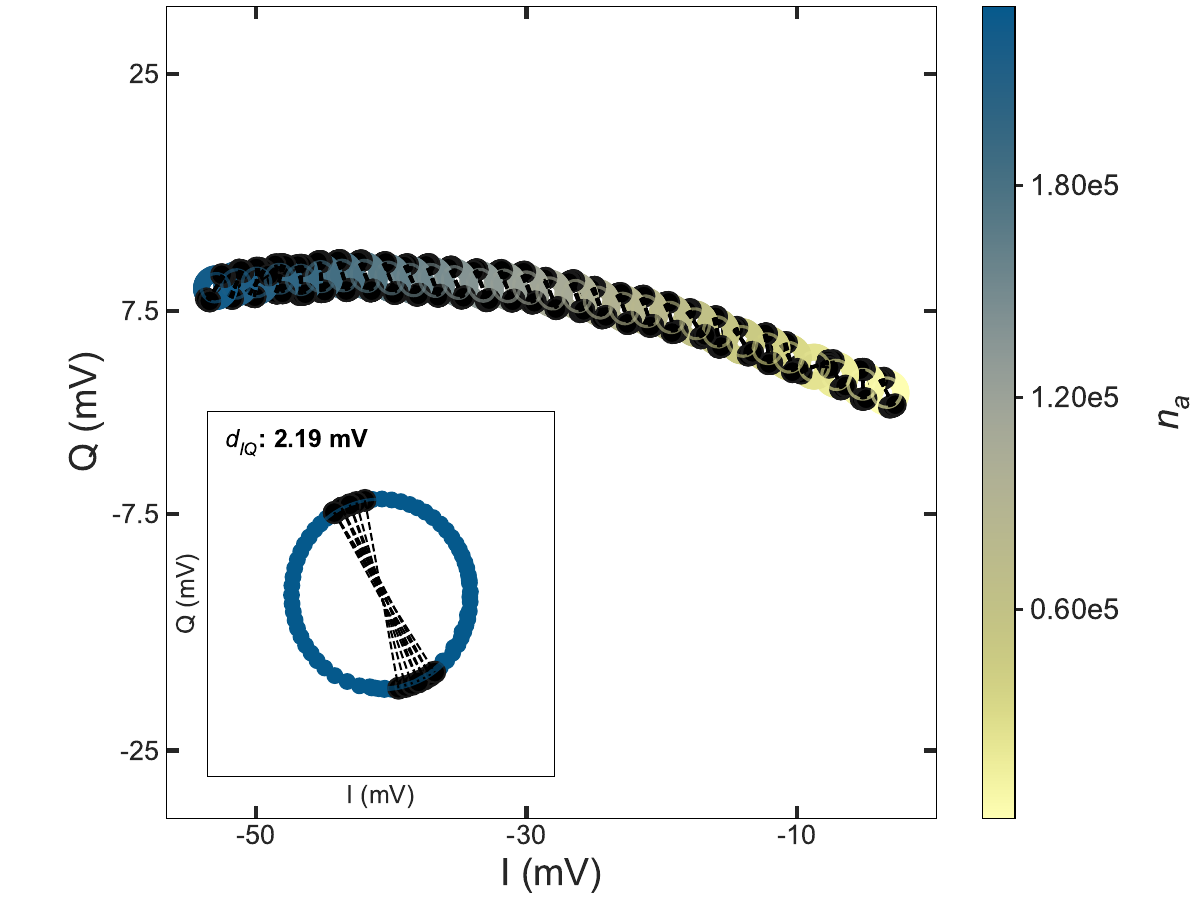}
    \caption{Illustration of the trajectory of the perturbative coherent signal in the IQ plane under Protocol A, as a function of the photon number inside the cavity $n_{a}$. (inset) A zoomed view of the perturbation for $n_{a} \sim 2.3 \times 10^5$ photon, where $d_{IQ}$ is calculated by averaging the maximum distances between different phase pairs.}
    \label{fig:na}
\end{figure}

In this section we determine the optimum working point of our detector, as the
spectral and coherence properties of the quantum circuit are expected to strongly depend on
$n_a$. Figure \ref{fig:na} illustrates the displacement in the IQ plane as the phase varies, for different photon numbers $n_a$ within the cavity. 

For each photon number $n_a$, a Ramsey pulse sequence with a total duration of $t_{sequence}=4 \mu s$, as defined in the figure \ref{fig:pulse_seq}, is generated by the AWG with a fixed number of traces $N$.  Initially, the cavity is populated with photons at its resonance frequency for $t_{pulse} = 2 \mu s$. Both timing and amplitude parameters were optimized for this measure, and the circuit photon number, $n_q$ remains fixed at $n_q = 3 \times 10^4$ photon. Typically, $t_{ramsey} = 250$ns and  $t_{pulse} = 30$ns. 

Simultaneously, a continuous tone from an RF source is applied to simulate a perturbative coherent signal, $\varphi_{\mathrm{w}}$. Its phase varies from $0$ to $2\pi$ using two phase shifters, resulting in the circular pattern in the I-Q plan. The y-axis value, $d_{IQ}$, in the main text is calculated by averaging the ten largest distances in the circular signature, as shown in the miniature. This procedure is repeated for each $n_{a}$ value, corresponding to different amplitudes of the cavity drive.

 \newpage

\begin{figure}[htbp]
    \centering
    \includegraphics[width=0.6\textwidth]{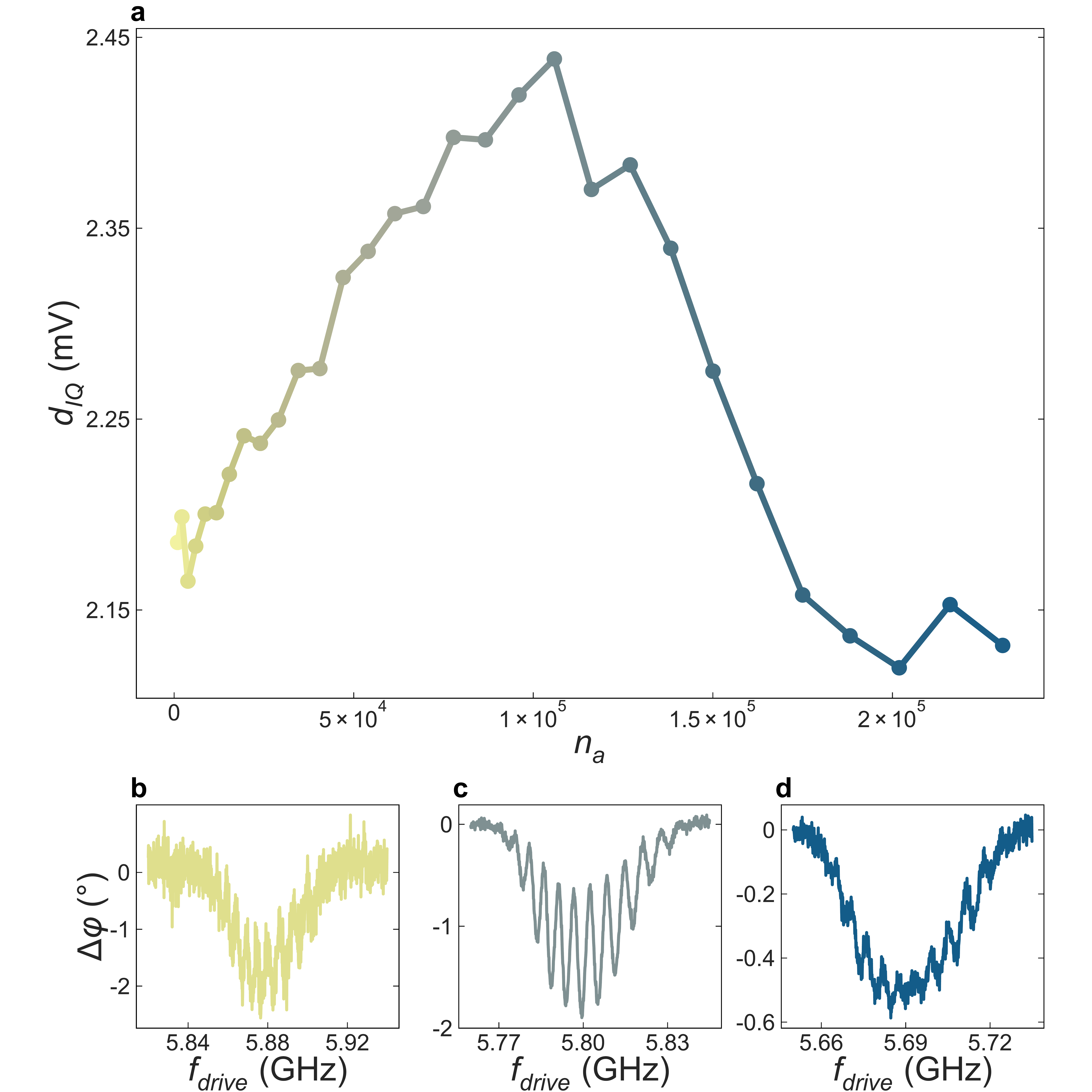}
    \caption{a) Illustration of the optimal pump power under Protocol A as a function of the cavity photon number $n_{a}$. $d_{IQ}$ denotes the maximum distance between two different phase values in the IQ plane. b), c), d) Corresponding Ramsey fringes for increasing values of $n_a$. At low photon numbers, the signal is dominated by noise; as $n_a$ increases, an optimum is reached, beyond which the Ramsey fringes are degraded, which could be due to measurement-induced dephasing.}
    \label{fig:fig3}
\end{figure}

To optimize the readout performance under Protocol A, we study the dependence of the measurement contrast on the cavity photon number $n_a$. As shown in \ref{fig:fig3}, the optimal pump power corresponds to a maximal separation $d_{IQ}$ between the circuit states in the IQ plane. At low photon numbers, the measurement signal is weak and dominated by noise. Increasing $n_a$ enhances the contrast up to an optimal value, beyond which it begins to deteriorate. This degradation is manifested by the disappearance of Ramsey fringes and may be attributed to measurement-induced dephasing. This trade-off establishes an optimal operating point for achieving maximum detection sensitivity.

\newpage
\subsection{Calibration of the linear amplifier sensitivity.}

\begin{figure}[htbp]
    \centering
    \includegraphics[width=0.6\textwidth]{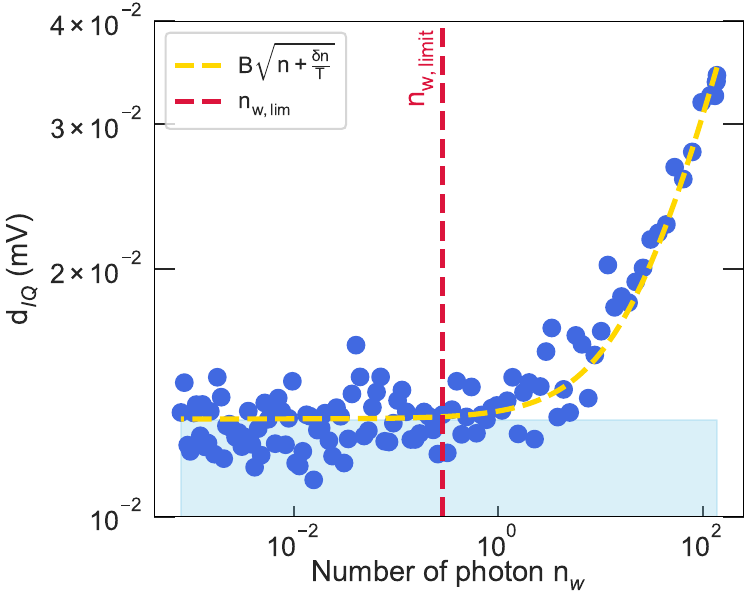}
    \caption{Calibration of the linear amplifier sensitivity, with $t_{meas} = 4\mu s$. The fit gives $\delta n = 1.21 \times 10^5$ and $B = 2.77 \times 10^{-3}$.}
    \label{fig:hemt}
\end{figure}

The sensitivity of the linear amplifier was calibrated by applying the previously described detection protocol to the cavity mode, without the circuit addition. The Ramsey sequence is performed far from the circuit frequency, at 5.5 GHz. The cavity photon number was set to the optimal value determined from Fig.~\ref{fig:fig3}, namely $10^{5}$ photons. The acquisition time corresponds to a single iteration of the complete pulse sequence, i.e $T = 4 \mu s$. Figure \ref{fig:hemt} illustrates this calibration procedure and reveals a sensitivity threshold around 10 photons, which is consistent with our linear amplifier noise specifications. The data are fitted using the model $d_{IQ} = B \sqrt{n_w + \frac{\delta n}{T}}$. The dashed line $n_{w,lim}$ corresponds to the sensitivity limit, which is determined following the criteria $\left| \frac{f-\sigma}{f} \right| < 5 \%$. 

\newpage
\subsection{Additional Data: Protocol A}

Additional Data supporting Fig. 4 from the main text are provided. The detection protocol is defined as in the previous section, except this time, the value of the photon number inside the cavity, $n_a$, and in the circuit, $n_q$ are optimized once and remain fixed. 

In parallel to the Ramsey sequence, the perturbative coherent signal phase, $\varphi_{\mathrm{w}}$, varies to get the circular pattern in the IQ-plan. The amplitude of the perturbation is again determined with the average of the largest distances in the circular signature, and is denoted as $d_{IQ, max}$. This procedure is repeated for each $n_{\mathrm{w}}$ value, corresponding to different levels of applied attenuation, which is added to the fixed attenuation of $\sim 93$ dB. 

The timing of the measure is defined as $ N \times 4 \mu s$. We find the optimal value to be $600 ms$, i.e. $1.5 \times 10^5$ traces. The noise $\sigma$ of each measurement is defined by the standard deviation of the most attenuated values. The data are fitted by the function $f(n_{\mathrm{w}}) = A \sqrt{n_{\mathrm{w}} + \frac{(\delta n_{\mathrm{w}})^2}{t}}$ with $ A \sim 2 \times 10^{-3}$, $\delta n_{\mathrm{w}} \sim 10^{-1}$,$n_{\mathrm{w}}$ the photon number in the coherent signal and $t$ the integration time. 

\begin{figure}[htbp]
    \centering
    \includegraphics[width=0.6\textwidth]{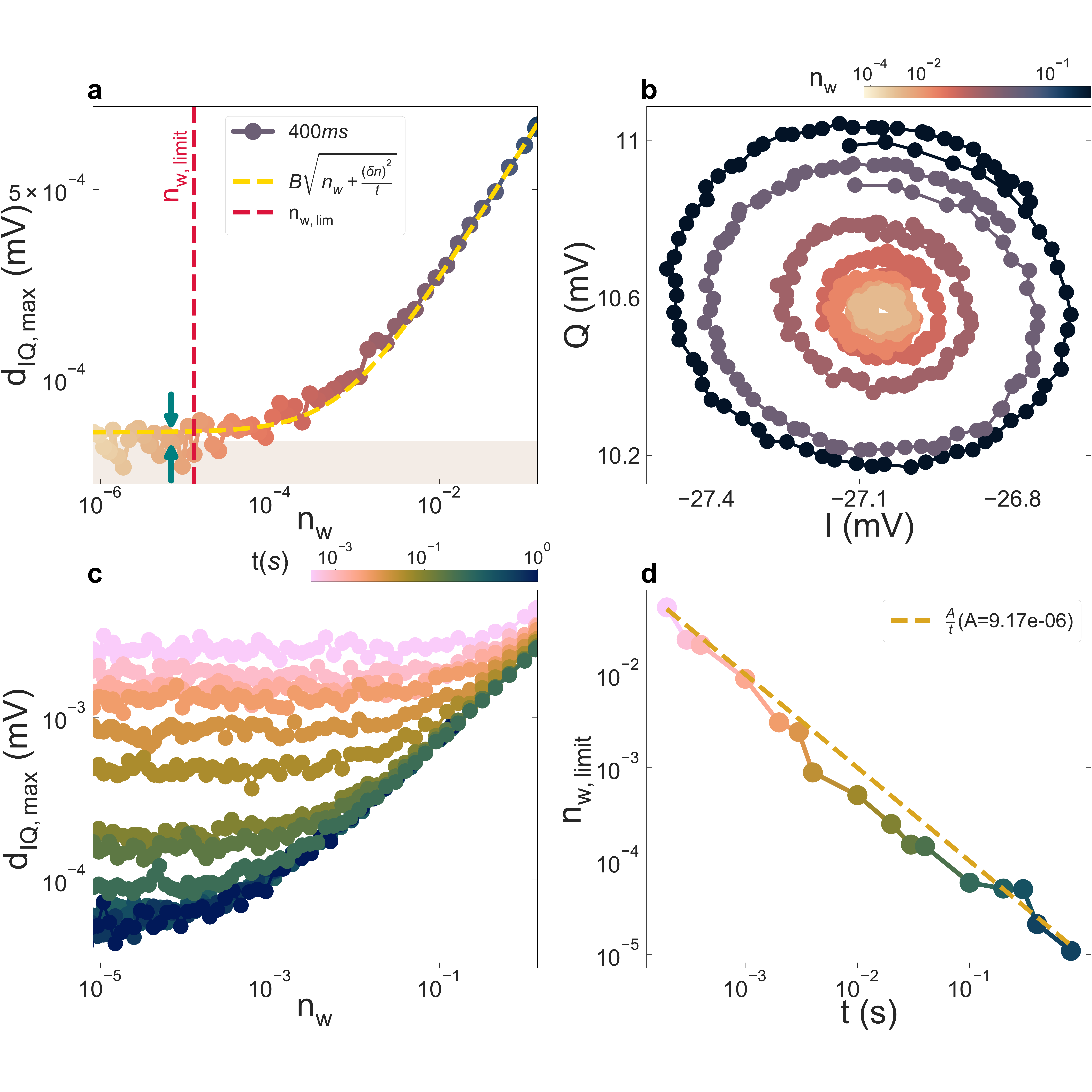}
    \caption{Details on Protocol A : a) Detection protocol A results for $t=400$ $ms$, with fitted results and fonction. The noise is depicted as the brown-shaded area, while two blue arrows indicate the convergence criteria previously defined. b) Illustration of the detection protocol for $t=400$ $ms$ in the IQ-plan. The perturbation circle signature weakens as $n_{\mathrm{w}}$ decreases, eventually merging with the noise level. c) Detection Protocol A for varying integration times, ranging from $t=100$ $\mu s$ to $t=800$ $ms$. d) Sensitivity as a function of integration time $t$ $(s)$, fitted by the model function $f(t)=\frac{A}{t}$. }
    \label{fig:S4_Protocole_a}
\end{figure}

The sensitivity of each experiment is determined by the criteria $\left| \frac{f-\sigma}{f} \right| < 5 \%$, and the sensitivity over time is plotted on Figure \ref{fig:S4_Protocole_a}d. 

\newpage

\subsection{Additional Data : Protocol B}

A second approach, referred to as Protocol B, employs the same pulse sequence as Protocol A: the cavity is driven at its resonance frequency while the superconducting circuit undergoes the Ramsey sequence. Simultaneously, a continuous tone representing the perturbative coherent signal is applied via an RF source, with its amplitude $n_{\mathrm{w}}$ and phase $\varphi_{\mathrm{w}}$ adjustable.  

The key difference lies in how the perturbation is measured, as it is now directly probed through the superconducting circuit's frequency.

This protocol begins with a full scan of all phases $\varphi_{\mathrm{\mathrm{w}}}$, from $0 $ to $2 \pi$. From this scan, the maximum and minimum phase values, $\varphi_{\mathrm{w},max}$  and  $\varphi_{\mathrm{w}, min}$, are extracted and stored. The difference between those two phases, representing the amplitude of the oscillations, is then evaluated as a function of $n_{\mathrm{w}}$, the photon number in the perturbative coherent signal. 

The noise level is determined as the standard deviation of the saturated values at the lowest $n_{\mathrm{w}}$. The data can be fitted using the function $g(n_{\mathrm{w}}) = A \sqrt{n_{\mathrm{w}}}$. One can note that, compared to Protocol A, this method is independent of measurement time since it relies solely on evaluating two phase values, and not every one of them.

\begin{figure}[htbp]
    \centering
    \includegraphics[width=0.6\textwidth]{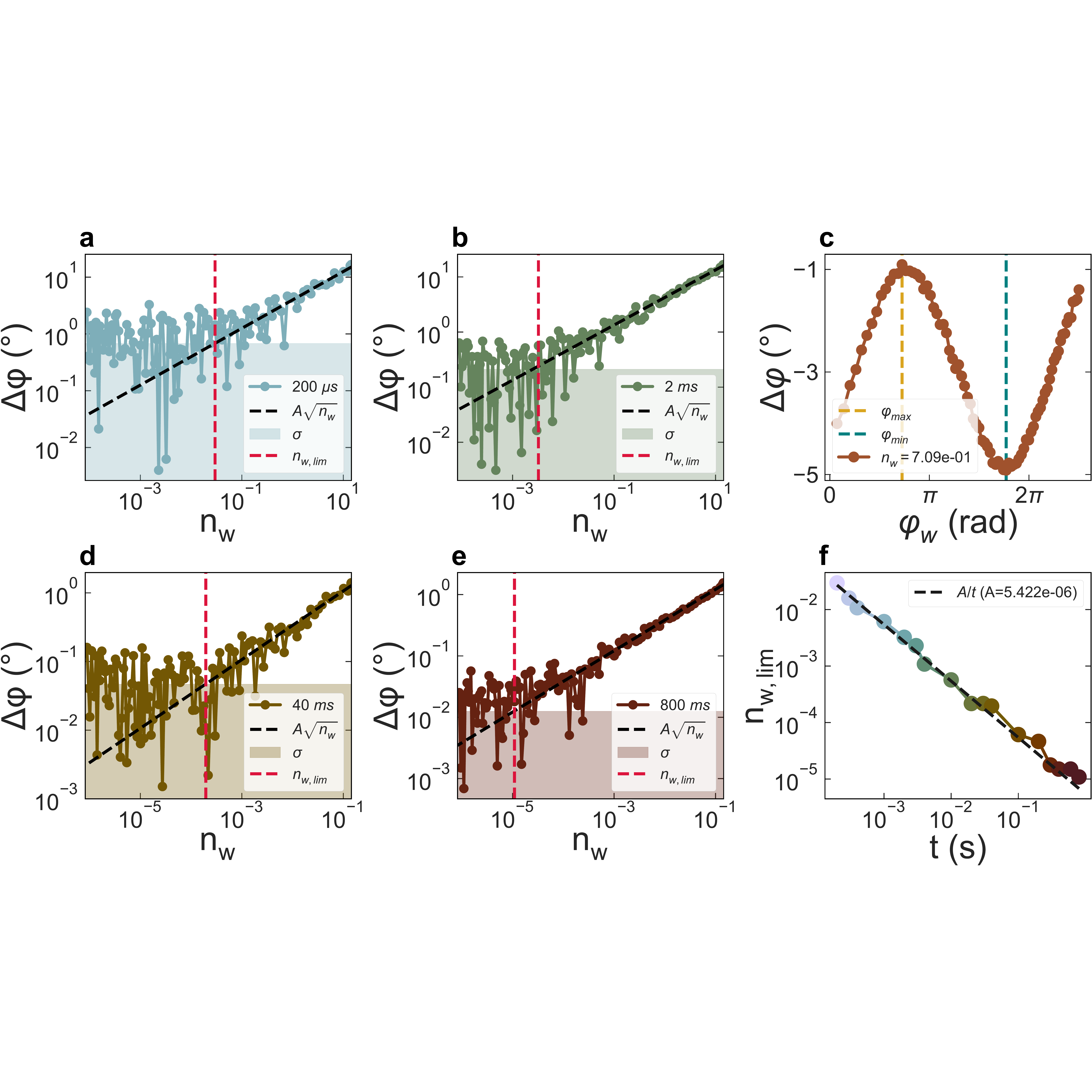}
    \caption{Details on Detection Protocol B : [a),b),d),e)] Detection Protocol B for $t={[200 \mu s, 2 ms, 40ms, 800ms]}$. The data are fitted by the function $g(n_{\mathrm{w}}) = A \sqrt{n_{\mathrm{w}}}$, and the baseline is indicated by the lighter-colored region. $n_{w,lim}$ is defined as the intersection between the baseline and the fit function. c) For $n_{\mathrm{w}} = 7.09 \times 10^{-1}$ , the figure illustrates the determination of two phases $\varphi_{w,max}$ and $\varphi_{w,min}$. 
    f) Sensitivity of detection protocol B : $n_{w,lim}$ is plotted as a function of the integration time $t(s)$. }
    \label{fig:S4_Protocole_b}
\end{figure}

The sensitivity for each experiment is determined by the intersection between the fitted function and the baseline, and is plotted on Figure \ref{fig:S4_Protocole_b}.f).

\newpage

\subsection{Influence of the detection frequency $f_{Ramsey}$ on the detection protocol}

\begin{figure}[htbp]
    \centering
    \includegraphics[width=0.6\textwidth]{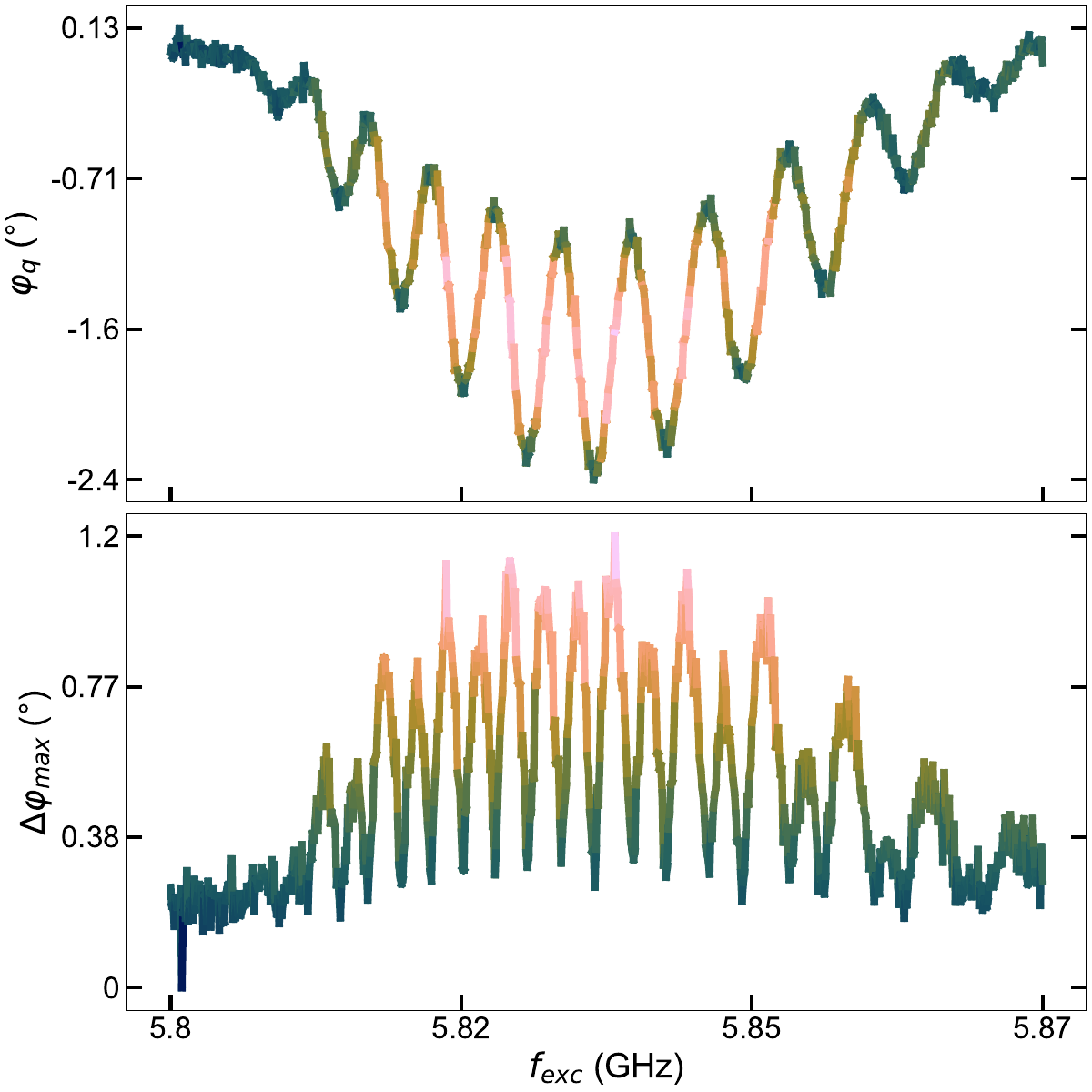}
    \caption{Influence of $f_{Ramsey}$ on the detection protocol B : Phase contrast $\varphi_q$ of the Ramsey fringes as a function of drive frequency $f_{exc}$ (top). Sensitivity of  the detection protocol B for each drive frequency, for $n_{\mathrm{w}} \sim 10^{-1}$ (bottom) }
    \label{fig:freq}
\end{figure}

In the main text, we have considered an optimized detection frequency at the middle of the right slope of the most contrasted Ramsey fringe. We can verify this by duplicating the protocol from Figure 4.c of the main text : we measure the phase difference $\Delta \varphi_{max}$ of the superconducting circuit at a fixed frequency for two wion phases : $0$ and $\pi$. Figure \ref{fig:freq} shows us that the best $\Delta \varphi_{max}$ is obtained when the Ramsey frequency is located at the highest slope, in agreement with other Ramsey interferometry experiments \cite{Degen}. 

Moreover, Figure \ref{fig:freq} shows that detection is more effective on the "external" slope compared to the "internal" slope, relative to the center of the Ramsey fringes.

\newpage
\subsection{Influence of the number of photons $n_{b}$ in the superconducting circuit on the Ramsey fringes}

\begin{figure}[!htbp]
    \centering
    \includegraphics[width=0.6\textwidth]{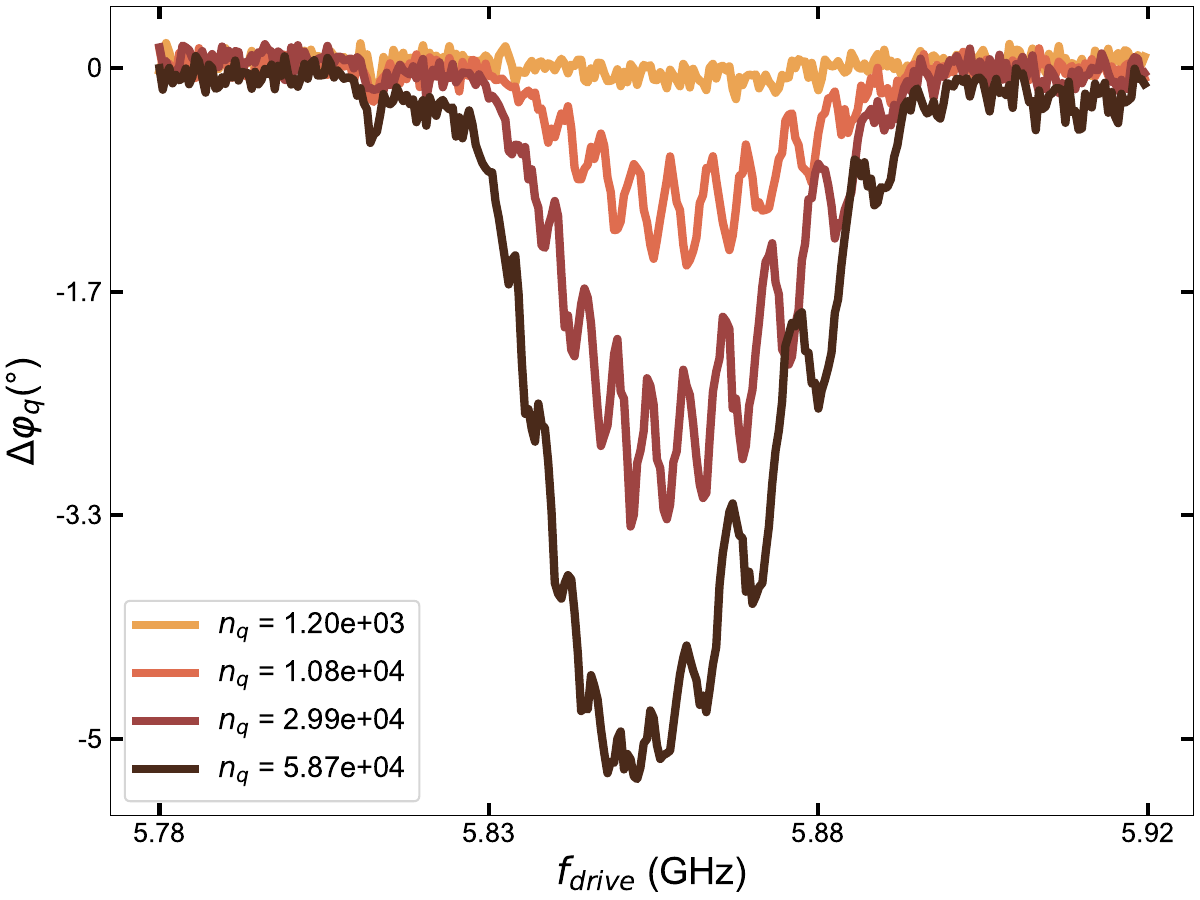}
    \caption{Phase contrast $\Delta \varphi_q$ of the Ramsey fringes as the function of drive frequency $f_{drive}$ for different $n_{q}$.}
    \label{fig:nq}
\end{figure}

Figure \ref{fig:nq} shows the influence of the number of photons $n_{q}$ in the circuit. We see that increasing the number of photons $n_{q}$ gives a better contrast, up to a certain point where we loose the Ramsey fringes completely. This is also a direct consequence of the measurement-induced dephasing, as seen in \cite{JGambetta}.

\subsection{Influence of the readout frequency $f_{rdt,cav}$ on the Ramsey fringes}

\begin{figure}[!htbp]
    \centering
    \includegraphics[width=0.6\textwidth]{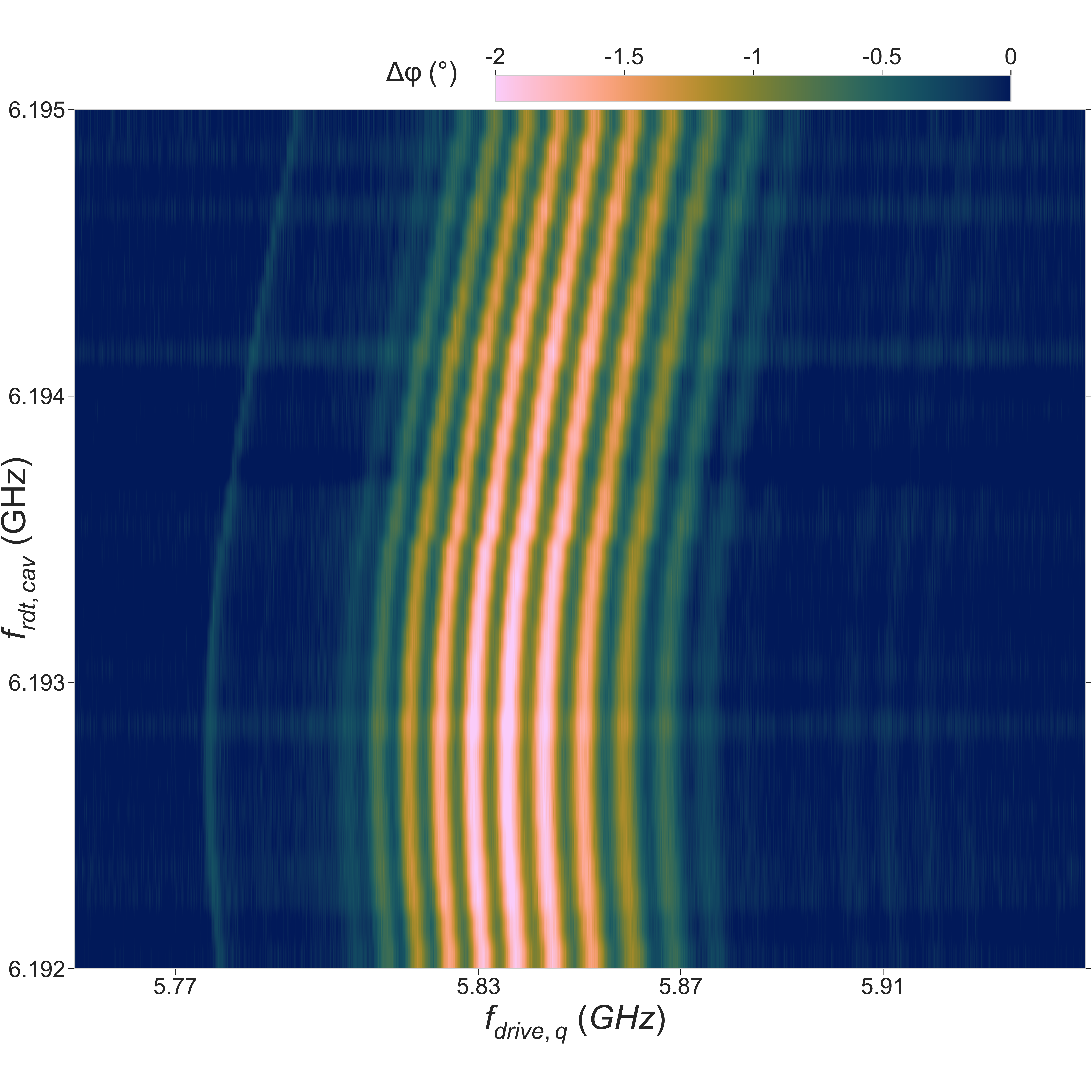}
    \caption{Phase contrast $\Delta \varphi$ of the Ramsey fringes as a function of the drive frequency $f_{drive,q}$ and the cavity readout frequency $f_{rdt,cav}$.}
    \label{fig:frdt}
\end{figure}

Figure \ref{fig:frdt} shows how the readout frequency $f_{rdt,cav}$ affects the Ramsey fringes. The optimal contrast is achieved with a frequency slightly shifted to the right of the cavity resonance, at 6.193 GHz. This can be explained by the non-linearity induced by the cavity.

\subsection{Influence of the Power $P_{m}$ on the detection protocol}
\begin{figure}[!htbp]
    \centering
    \includegraphics[width=0.6\textwidth]{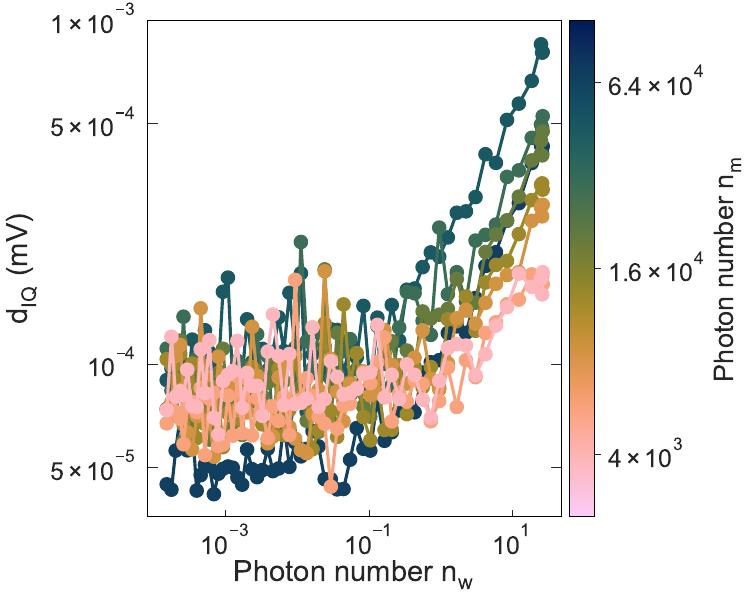}
    \caption{Effect of the drive power $P_m$ on the detected signal slope and its relation to the detection protocol. The optimal value of $P_m$ is consistent with the results presented in Figure 4 of the main text.}
    \label{fig:Pm}
\end{figure}

Figure \ref{fig:Pm} illustrates the influence of the power $P_{m}$ on the detection protocol discussed in the last section of the main text. The slope of the detected signal is directly affected by the amplitude of the drive $P_m$ applied to the auxiliary mode. An optimal value is observed, which aligns with the findings presented in Figure 4 of the main text.

\end{document}